\documentclass[preprint,12pt,sort&compress]{elsarticle}

\usepackage{amssymb}
\usepackage[version=4]{mhchem}
\usepackage{siunitx}

\usepackage{tikz}
\usetikzlibrary{arrows.meta,patterns,shapes.arrows}
\usetikzlibrary{ipe} 
\tikzstyle{ipe stylesheet} = [
  ipe import,
  even odd rule,
  line join=round,
  line cap=butt,
  ipe pen normal/.style={line width=0.4},
  ipe pen heavier/.style={line width=0.8},
  ipe pen fat/.style={line width=1.2},
  ipe pen ultrafat/.style={line width=2},
  ipe pen normal,
  ipe mark normal/.style={ipe mark scale=3},
  ipe mark large/.style={ipe mark scale=5},
  ipe mark small/.style={ipe mark scale=2},
  ipe mark tiny/.style={ipe mark scale=1.1},
  ipe mark normal,
  /pgf/arrow keys/.cd,
  ipe arrow normal/.style={scale=7},
  ipe arrow large/.style={scale=10},
  ipe arrow small/.style={scale=5},
  ipe arrow tiny/.style={scale=3},
  ipe arrow normal,
  /tikz/.cd,
  ipe arrows, 
  <->/.tip = ipe normal,
  ipe dash normal/.style={dash pattern=},
  ipe dash dotted/.style={dash pattern=on 1bp off 3bp},
  ipe dash dashed/.style={dash pattern=on 4bp off 4bp},
  ipe dash dash dotted/.style={dash pattern=on 4bp off 2bp on 1bp off 2bp},
  ipe dash dash dot dotted/.style={dash pattern=on 4bp off 2bp on 1bp off 2bp on 1bp off 2bp},
  ipe dash normal,
  ipe node/.append style={font=\normalsize},
  ipe stretch normal/.style={ipe node stretch=1},
  ipe stretch normal,
  ipe opacity 10/.style={opacity=0.1},
  ipe opacity 30/.style={opacity=0.3},
  ipe opacity 50/.style={opacity=0.5},
  ipe opacity 75/.style={opacity=0.75},
  ipe opacity opaque/.style={opacity=1},
  ipe opacity opaque,
]

\usepackage{caption}
\usepackage{subfig}
\usepackage{floatrow}

\usepackage[hidelinks]{hyperref}

\usepackage{mathtools}

\usepackage{booktabs, tabularx}

\journal{Electrochimica Acta}

\begin{document}

\begin{frontmatter}



\title{Electrolyte effects on the alkaline hydrogen evolution reaction: a mean-field approach}


\author[label1]{Lucas B.T. de Kam\corref{cor1}\fnref{fn1}}
\ead{l.b.t.de.kam@lic.leidenuniv.nl}
\cortext[cor1]{Corresponding authors}
\fntext[fn1]{Present address: Leiden Institute of Chemistry, Einsteinweg 55, 2333CC Leiden, The Netherlands}

\author[label1]{Thomas L. Maier}
\author[label1]{Katharina Krischer\corref{cor1}}
\ead{krischer@tum.de}

\affiliation[label1]{organization={Nonequilibrium Chemical Physics, Department of Physics, Technische Universität München},
            addressline={James-Franck-Straße 1}, 
            city={Garching bei München},
            postcode={85748}, 
            country={Germany}}

\begin{abstract}
This paper introduces the combination of an advanced double-layer model with electrochemical kinetics to explain electrolyte effects on the alkaline hydrogen evolution reaction. It is known from experimental studies that the alkaline hydrogen evolution current shows a strong dependence on the concentration and identity of cations in the electrolyte, but is independent of pH. To explain these effects, we formulate the faradaic current in terms of the electric potential in the double layer, which is calculated using a mean-field model that takes into account the cation and anion sizes as well as the electric dipole moment of water molecules. We propose that the Volmer step consists of two activated processes: a water reduction sub-step, and a sub-step in which \ce{OH^-} is transferred away from the reaction plane through the double layer. Either of these sub-steps may limit the rate. The proposed models for these sub-steps qualitatively explain experimental observations, including cation effects, pH-independence, and the trend reversal between gold and platinum electrodes. We also assess the quantitative accuracy of the water-reduction-limited current model; we suggest that the predicted functional relationship is valid as long as the hydrogen bonding structure of water near the electrode is sufficiently maintained.
\end{abstract}



\begin{keyword}
hydrogen evolution reaction \sep cation effect \sep electric double layer \sep water structure \sep electrochemical kinetics



\end{keyword}

\end{frontmatter}


\section{Introduction}
The rate of electrochemical reactions is intimately linked to the structure of the double layer \citep{delahay1965, ledezma2017interfacial, sarabia2018effect, xue2018influence, monteiro2021absence, monteiro2021understanding, goyal2021interrelated, bender2022understanding, ringe2023cation, li2023electric}. The common understanding of the relation between the double layer and kinetics is still based on Frumkin's idea. In his work \citep{frumkin1933wasserstoffuberspannung}, Frumkin relates the current to the potential in the reaction plane, which he calculates using the Gouy-Chapman-Stern double-layer model \citep{BardFaulkner2001}. Over the last couple of decades, continuum models of the double layer have advanced significantly \citep{borukhov2000adsorption, abrashkin2007dipolar, iglivc2019differential, huang2021hybrid}, achieving reasonable success in describing the double-layer capacitance \citep{landstorfer2022thermodynamic, huang2023density, deKam2023}. Recently, studies have explored the usage of such models in the context of kinetics for oxygen evolution \citep{huang2021cation} and carbon dioxide reduction \citep{ringe2019understanding}. Here, we focus on the alkaline hydrogen evolution reaction. Its comparatively simple reaction mechanism allows us to develop an intuitive model without relying on quantum chemical calculations. By combining Frumkin's approach with the double-layer model of Igli\v{c} et al. \citep{iglivc2019differential}, we explain experimentally observed electrolyte effects. In addition, where previous studies mostly make qualitative comparisons with experimental data \citep{ringe2019understanding, huang2021cation}, we identify the regime in which our model demonstrates the correct functional relationship.

Let us first review the mechanism of hydrogen evolution in alkaline media on metal electrodes. The reaction mechanism consists of two consecutive elementary steps. The first one is the so-called Volmer step, where a hydrogen atom is adsorbed on the catalyst surface:
\begin{equation} \label{eq:volmer}
    \ce{H_2O + e^- + $*$ <=> H^* + OH^-},
\end{equation}
with $*$ denoting a free adsorption site on the catalyst. In the literature, it is usually understood that the hydroxide ion (\ce{OH^-}) on the right-hand side is located outside the double layer, where electroneutrality holds. The Volmer step may be followed by the Tafel step, where adsorbed hydrogen atoms combine to form molecular hydrogen:
\begin{equation}
    \ce{2 H^* <=> H_2 + 2$*$},
\end{equation}
or by the Heyrovsky step, in which water molecules react with adsorbed hydrogen atoms to form molecular hydrogen and hydroxide ions:
\begin{equation}
    \ce{H^* + H_2O + e^- <=> H_2 + OH^- + $*$}.
\end{equation}
The overall reaction rate is usually limited by the slow water reduction steps, i.e. the Volmer step \citep{mccrum2020role, qin2023cation, rebollar2018determining, monteiro2021understanding} or the Heyrovsky step \citep{rebollar2018determining, monteiro2021understanding}. 

The alkaline hydrogen evolution current shows a strong dependence on the concentration and identity of cations in the electrolyte. At moderately alkaline pH, increasing the cation bulk concentration yields a larger current \citep{goyal2021interrelated}. The current is also larger for more weakly hydrated cation species \citep{bender2022understanding, xue2018influence, goyal2021understanding}. Various studies explain these effects in terms of the cation concentration in the double layer \citep{monteiro2021understanding, ringe2023cation}. The interfacial cation concentration is presumed to be larger for larger bulk concentrations and more weakly hydrated species, which pack more tightly in the double layer \citep{borukhov2000adsorption}. For example, it is smaller in \ce{Li^+}-based electrolytes than in \ce{Na^+}-based electrolytes. The model of \citet{koper2023theory} suggests that the presence of cations in the double layer decreases the activation barrier of the water reduction step by changing the electric potential at the reaction plane, akin to Frumkin's idea. However, \citeauthor{koper2023theory} did not compute this electric potential, and could therefore not compare the theory to experimental data.

At strongly alkaline pH, increasing the cation concentration starts to have an adverse effect \citep{monteiro2021understanding}. This turning point occurs at a lower pH for platinum electrodes than for gold \citep{monteiro2021understanding}, and the observed cation trends are reversed between gold and platinum at pH 13 \citep{bender2022understanding, xue2018influence}. \citet{bender2022understanding} rationalize the trend reversal by splitting the Volmer step into two sub-steps: a water reduction sub-step,
\begin{equation}
    \label{eqn:sub1}
    \ce{H_2O + e^- + $*$ <=> H^* + OH^- (reaction plane)},
\end{equation}
and a sub-step in which \ce{OH^-} is transferred away from the reaction plane through the double layer, 
\begin{equation}
    \label{eqn:sub2}
    \ce{OH^- (reaction plane) <=> OH^- (bulk)}
\end{equation}
where by `bulk' the electroneutral electrolyte outside the double layer is implied. Considering strongly alkaline conditions, \citeauthor{bender2022understanding} propose that for gold, the reaction rate is limited by the water reduction sub-step \eqref{eqn:sub1}, whereas for platinum, the rate is limited by the \ce{OH^-} transfer sub-step \eqref{eqn:sub2}.

In acidic media, the activity of the hydrogen evolution reaction is independent of pH on the RHE scale (and thereby independent of the interfacial electric field) \citep{briega2020activity}, corresponding to a Nernstian shift of $-\qty{59}{mV/pH}$ on the SHE scale. In contrast, alkaline hydrogen evolution shows pH-independence on the SHE scale under conditions where the water reduction sub-step is rate-limiting \citep{strmcnik2013improving, ringe2023cation}, suggesting that the interfacial electric field plays an important role.
Under conditions where it is proposed that \ce{OH^-} transfer hampers the rate, the current again depends on pH \citep{strmcnik2013improving, goyal2021interrelated, ringe2023cation}. 

In this work, we develop expressions for the water-reduction-limited current and the \ce{OH^-}-transfer-limited current in terms of the electric potential in the double layer, which is computed from the mean-field model of Igli\v{c} et al. \citep{iglivc2019differential}. The model for the water-reduction-limited current explains the pH-independence and cation trends on gold electrodes and quantitatively describes the data in \ce{Li^+}-based electrolytes. The simple model we propose for the \ce{OH^-}-transfer-limited current considers the transfer as another activated process. The model predicts reversed parameter trends for the water-reduction-limited and the \ce{OH^-}-transfer-limited currents, the former prediction being consistent with Au measurements, and the latter with Pt data. In our analysis, we focus on the Volmer step, but a similar approach can be taken for the Heyrovsky step.

\section{Methods}
In this section, we first summarize the double-layer model that we will use to calculate the electric potential in the double layer. We then connect the electric potential to the electrode potentials used in electrochemical kinetics and derive expressions for when either the water reduction substep or the \ce{OH^-} transfer substep is rate-determining.

\begin{figure}
    \centering
    \includegraphics[width=3.248in]{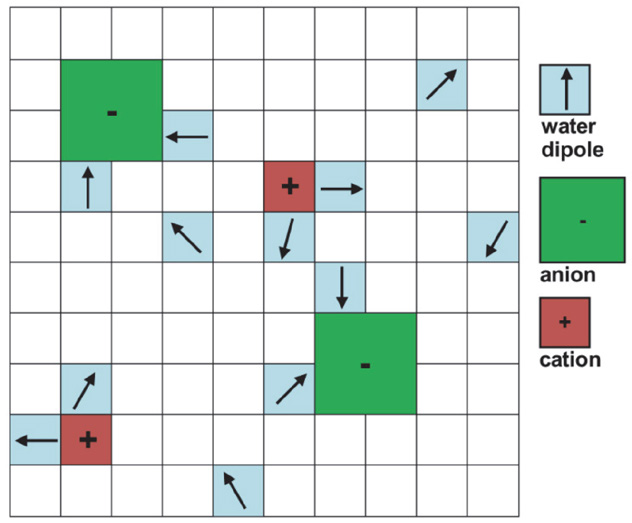}
    \caption{Water dipoles, anions, and cations on a grid. Note that the grid considered in the theory is in fact three-dimensional. Here, the specific case $\gamma_+=1$ and $\gamma_-=2^3$ is shown. $\gamma_\pm$ is defined in Eq. \eqref{eq:gamma} in the main text. In this work, we consider the case that there are no lattice vacancies. Reprinted from Igli\v{c} et al. \citep{iglivc2019differential} under a CC-BY 4.0 license.}
    \label{fig:lattice-dipoles}
\end{figure}

\subsection{Double-layer model} \label{sec:edl-model}
In the model of Igli\v{c} et al. \citep{iglivc2019differential}, ions and dipolar water molecules are placed on a three-dimensional lattice (Figure \ref{fig:lattice-dipoles}). The lattice spacing $a$ is defined such that when all lattice sites are filled by water molecules, their concentration equals the number density of pure water, \qty{55.5}{M}. This gives $a \approx \qty{0.31}{nm}$.

Ion species may occupy several sites. The effective size $\gamma$ of an ion is the number of sites occupied by an ion and its hydration shell,
\begin{equation} \label{eq:gamma}
    \gamma_\pm = \left( \frac{d_\pm}{a} \right)^3
\end{equation}
where $d_\pm$ are the diameters of the solvated cat- or anions. Typically, ions with a small radius such as \ce{Li^+} have a larger hydration shell than ions with a large radius such as \ce{Cs^+} \citep{tielrooij2010cooperativity}. Based on the hydration number of ions \citep{israelachvili2015intermolecular} and estimated ion sizes from dielectric decrement \citep{hasted1948dielectric}, we choose $\gamma_+=4,5,6,7$ to correspond roughly to \ce{Cs^+}, \ce{K^+}, \ce{Na^+} and \ce{Li^+}. We also choose $\gamma_-=2$ for hydroxide ions and other anions, as anions are typically more weakly hydrated \citep{hasted1948dielectric, israelachvili2015intermolecular}. 

In the mean-field approximation, it can be shown \citep{borukhov2000adsorption, abrashkin2007dipolar, huang2021hybrid} that the electric potential $\phi$ extending into the electrolyte is described by the one-dimensional Poisson's equation,
\begin{equation} \label{eq:pbe}
    -\frac{\partial}{\partial x} \left( \varepsilon \frac{\partial \phi}{\partial x} \right) = e_0 (n_+ - n_-).
\end{equation}
Here, $e_0$ is the elementary charge, and $\varepsilon$ is the dielectric permittivity, which depends non-linearly on the local electric field $E=-\partial \phi / \partial x$ due to the polarization of dipolar water molecules \citep{booth1951dielectric}:
\begin{equation} \label{eq:permittivity}
    \varepsilon = \varepsilon_\infty + \frac{n_\mathrm{w} p}{E} \mathcal L(\beta p E),
\end{equation}
where $\varepsilon_\infty\approx1.8$ is the optical permittivity of water \citep{israelachvili2015intermolecular}, $\beta=1/k_\mathrm{B} T$ the inverse temperature, and $\mathcal{L}(z) = \coth (z) - 1/z$ the Langevin function. The effective dipole moment of water molecules, $p$, is fitted such that the permittivity attains the value for pure water at $E=0$. $n_\pm$ are the cation and anion number densities, and $n_\mathrm{w}$ is the number density of water molecules; the superscript $*$ indicates bulk values. The ionic and water number densities depend on the local electric potential and electric field as \citep{iglivc2019differential}
\begin{align} \label{eq:gi-num-densities}
\begin{split}
    n_\pm &= n_\pm^* \frac{e^{\mp \beta e_0 \phi}} {\chi_\mathrm{w}^* \frac{\sinh \beta p E}{\beta p E} + \gamma_+ \chi_+^* e^{-\beta e_0 \phi} + \gamma_- \chi_-^* e^{\beta e_0 \phi}} \\
   n_\mathrm{w} &= n_\mathrm{w}^* \frac{\frac{\sinh \beta p E}{\beta p E}} {\chi_\mathrm{w}^* \frac{\sinh \beta p E}{\beta p E} + \gamma_+ \chi_+^* e^{-\beta e_0 \phi} + \gamma_- \chi_-^* e^{\beta e_0 \phi}}
\end{split}
\end{align}
where $\chi_\pm^* = n_\pm^*a^3$ the dimensionless bulk number densities, and $\chi_\mathrm{w}^* = 1 - \gamma_+\chi_+^* - \gamma_-\chi_-^*$. 

We only consider alkaline electrolytes with monovalent ions. The cation bulk concentration $c_+^*$ is a parameter that can be chosen to match experimental conditions. The pH determines the bulk concentration of hydroxide anions: $c_\mathrm{OH^-}^*=10^{-14+\mathrm{pH}} \;\mathrm{M}$. To vary the electrolyte pH and cation concentration independently, we add anions of some other monovalent species such that the total bulk concentration of anions $c_-^*$ equals the bulk cation concentration $c_+^*$, as required by electroneutrality. Hence, in this model, the pH does not affect the structure of the double layer. Number densities and concentrations are related as $n = N_\mathrm{A} c$, with $N_\mathrm{A}$ Avogadro's number. 

Igli\v{c} et al. \citep{iglivc2019differential} specify the boundary condition of the Poisson equation in terms of the surface charge. To apply the model in the context of kinetics, we need to specify the boundary condition in terms of the potential at the electrode, $\phi_0$. To do this, we need to realize that Equation \eqref{eq:pbe} describes the potential in the electrolyte only up to the plane of closest approach for electrolyte ions, located at $x=x_2$. Between the electrode surface at $x=0$ and $x=x_2$ there are no charges, so the potential profile is linear. Given $\phi_0$ and the potential at $x_2$, $\phi_2$, the boundary condition at $x_2$ reads
\begin{equation} \label{eq:stern-bc}
    \phi_2 - \phi_0 = -E(x_2) x_2.
\end{equation}
In this work, we only consider negative electrode charges. Due to the strong electric forces, only cations approach the electrode surface. Hence, $x_2=d_+/2$. The boundary condition in the bulk is $\phi = 0$ by choice of potential reference. The numerical implementation to solve the double-layer model (Eq. \ref{eq:pbe}) with its boundary condition (Eq. \ref{eq:stern-bc}) is discussed in \ref{app:numerical}.

The mean-field model captures several features of the double layer that the Gouy-Chapman-Stern model ignores. First, ion concentrations attain a saturation value of $n_\pm \to 1/(a^3 \gamma_\pm)$ as $\exp(\mp \beta e_0 \phi) \to \infty$. Second, the permittivity $\varepsilon$ decreases as the electric field $E$ increases, down to a limiting value of $\varepsilon_\infty$. This phenomenon is known as dielectric saturation \citep{booth1951dielectric}. However, the model ignores the hydrogen bonding structure of water. Besides, solvation shells are highly dynamic and complex \citep{tielrooij2010cooperativity, alfarano2021stripping}, but are only captured by a size parameter here. 

\subsection{Electrode potential and potential of zero charge} \label{sec:potentials}
The double-layer model computes the electric potential $\phi(x)$ in the electrolyte, given an electric potential $\phi_0$ applied at the electrode. However, in electrochemical kinetics, we usually deal with the electrode potential $\mathsf{E}$. To connect double-layer models to kinetics, we need to relate $\phi_0$ to $\mathsf{E}$. This connection is made by noting that the electrode potential is essentially the electrochemical potential of electrons in the electrode \citep{boettcher2020potentially},
\begin{equation}
    \mathsf{E} = -\frac{\tilde \mu_\mathrm{e}}{e_0} = -\frac{\mu_\mathrm{e}}{e_0} + \phi_0,
\end{equation}
where $\tilde \mu$ denotes an electrochemical potential and $\mu$ a chemical potential. Note again that the zero of the electric potential is chosen in the electrolyte bulk. At the point of zero charge (pzc), $\phi_0=0$, and so\footnote{In this discussion we neglect any metal surface potentials \citep{Schmickler2010} for simplicity. Simulations of \citet{huang2021grand} suggest that the surface potential of the metal does not depend strongly on the applied potential, and so this simplification does not qualitatively affect our results.}
\begin{equation} \label{eq:pzc}
    \mathsf{E}_\mathrm{pzc} = -\frac{\mu_\mathrm{e}}{e_0}.
\end{equation}
Hence, we may also write
\begin{equation} \label{eq:pzc-phi0}
    \mathsf{E} = \mathsf{E}_\mathrm{pzc} + \phi_0.
\end{equation}

To convert between $\mathsf{E}$ and $\phi_0$, we use values of $\mathsf{E}_\mathrm{pzc}$ measured experimentally. Because specific adsorption is absent in our model, the pzc defined above corresponds to the minimum in the differential capacitance, also known as the point of zero free charge (pzfc) \citep{rizo2015towards}. The pzfc is also closely related to the potential of maximum entropy (pme), which can be measured by laser-induced temperature jump methods \citep{climent2002potential}.

The different crystal orientations of gold and platinum have different work functions, and thus different potentials of zero charge. For the pzfc of Au(111), the capacitance minimum is located at \qty{0.51}{V} vs. SHE in a pH 3 electrolyte, according to \citet{ojha2020double}. For Pt(111) in acidic media, capacitance measurements \citep{ojha2020double}, calculations from CO displacement \citep{rizo2015towards}, and the laser-induced temperature jump method \citep{sebastian2017study} all yield a pzfc of around \qty{0.30}{V} vs SHE. Capacitance minimum data for Pt(100) and Pt(110) is (to our knowledge) not available, so we turn to measurements of the pme, which was determined to be at \qty{0.27}{V} and \qty{0.08}{V} vs. SHE at pH 1, respectively \citep{garcia2009potential}. However, the pme shows a significant dependence on pH: at pH 13, Pt(111), Pt(100) and Pt(110) have a pme of \qty{-0.09}{V}, \qty{-0.37}{V} and \qty{-0.54}{V} vs. SHE, respectively \citep{sarabia2020new}. In addition, the pme changes based on the cation species in the electrolyte \citep{briega2021cation}.

We will compare our model to experimental data obtained using Au(111), polycrystalline gold, and polycrystalline platinum electrodes, all in alkaline media. Based on the available data summarized above, assigning a pzc to these electrodes is a difficult task, particularly for the polycrystalline electrodes. However, even with the possibly large negative shift of the pzc in alkaline media, hydrogen always evolves at potentials negative of the pzc \citep{sarabia2020new}. In Section \ref{sec:double-layer-structure} we will show that the exact position of the pzc does not affect the results qualitatively, as long as the potential range under consideration is negative of the pzc. With this idea in mind, we for now choose $\mathsf{E}_\mathrm{pzc,Au}=\qty{0.51}{V}$ vs SHE for the Au(111) and polycrystalline gold electrodes and $\mathsf{E}_\mathrm{pzc,Pt}=\qty{0.30}{V}$ vs SHE for the polycrystalline platinum electrodes. Results for different choices of the pzc are shown in \ref{app:pzc}.

Above, we denoted electrode potentials versus the standard hydrogen electrode (SHE). When comparing to experimental data, we will sometimes denote electrode potentials versus the reversible hydrogen electrode (RHE), which depends on the electrolyte pH. One can convert between the scales as
\begin{equation} \label{eq:rhe-she-potential-conversion}
    \mathsf E \text{(vs. RHE)} = \mathsf E \text{(vs. SHE)} + \qty{59}{mV} \times \mathrm{pH}.
\end{equation}

\subsection{Kinetics}
Now that we have a way to calculate the electric potential in the double layer given the electrode potential, we derive expressions for the hydrogen evolution current density in terms of the electric potential. 

We first derive an expression for the current density in terms of the electric potential when the water reduction sub-step is rate-determining. Next, we consider the case where the transfer of \ce{OH^-} from within the double layer to electrolyte bulk limits the rate.

Note that the double-layer model describes the system in equilibrium, whereas theories of kinetics describe non-equilibrium situations. However, according to \citet{delahay1965}, the double-layer structure is not changed much at low current densities. We use this approximation here to explore the connection between the double-layer structure and kinetics.

\subsubsection{Water reduction} \label{sec:electron-transfer}
In the water reduction sub-step \eqref{eqn:sub1}, an electron is transferred from the electrode to a water molecule near the electrode surface. The water molecule thereby splits into an adsorbed hydrogen atom and a solvated hydroxide ion. We assume that the oxygen atom of the water molecule remains fixed in a position $x'\approx\qty{0.28}{nm}$ \citep{schmickler1993dependence} near the electrode surface. 

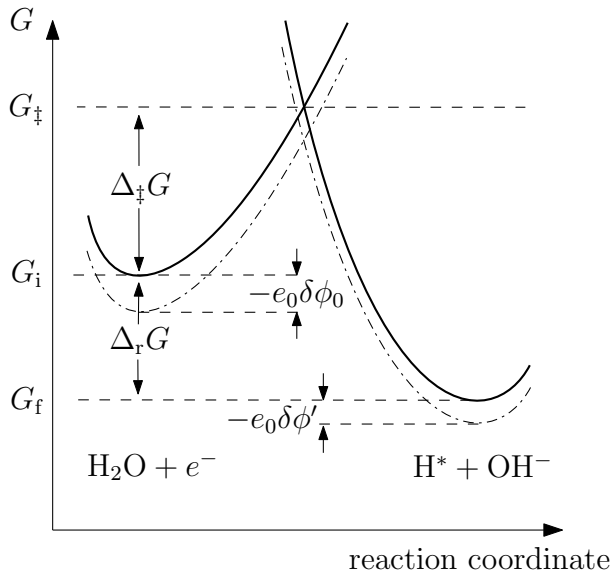
\begin{figure}
    \centering
    \begin{tikzpicture}[ipe stylesheet]
  \draw[shift={(125.753, 655.929)}, xscale=0.4288, yscale=0.5003, ipe dash dashed]
    (0, 0) rectangle (0, 0);
  \draw[shift={(64, 575.879)}, scale=1.0006, ->]
    (0, 0)
     -- (192, 0);
  \draw[shift={(64, 575.879)}, scale=1.0006, ->]
    (0, 0)
     -- (0, 192);
  \node[ipe node]
     at (175.659, 561.23) {reaction coordinate
};
  \node[ipe node]
     at (47.704, 764.399) {$G$};
  \node[ipe node]
     at (48.207, 669.407) {$G_\mathrm{i}$};
  \node[ipe node]
     at (48.011, 621.629) {$G_\mathrm{f}$
};
  \node[ipe node]
     at (47.392, 732.032) {$G_\ddag$
};
  \node[ipe node]
     at (85.484, 702.175) {$\Delta_\ddag G$
};
  \draw[ipe dash dashed]
    (73.6265, 624.988)
     -- (223.965, 624.828);
  \draw[-{>[ipe arrow small]}]
    (96.5334, 713.2103)
     -- (96.4854, 732.2323);
  \draw[-{>[ipe arrow small]}]
    (96.496, 697.2672)
     -- (96.552, 673.6942);
  \draw[ipe dash dashed]
    (156.326, 735.339)
     -- (73.1462, 735.461);
  \draw[shift={(158.408, 735.339)}, scale=0.4286, ipe dash dashed]
    (0, 0)
     -- (195.346, 0);
  \draw[-{>[ipe arrow small]}]
    (96.2932, 653.4754)
     -- (96.4612, 669.1004);
  \draw[-{>[ipe arrow small]}]
    (96.0982, 640.1844)
     -- (96.1902, 627.2164);
  \node[ipe node]
     at (85.38, 644.058) {$\Delta_\mathrm{r} G$
};
  \node[ipe node]
     at (76.966, 596.417) {$\mathrm{H_2O} + e^-$};
  \node[ipe node]
     at (199.703, 595.878) {$\mathrm{H^*} + \mathrm{OH^-}$};
  \draw[ipe pen heavier]
    (77.3644, 694.513)
     .. controls (82.4301, 670.343) and (96.1105, 666.755) .. (112.8813, 677.729)
     .. controls (129.652, 688.703) and (149.513, 714.239) .. (175.184, 767.202);
  \draw[ipe pen heavier]
    (151.675, 768.013)
     .. controls (182.48, 609.934) and (230.039, 611.826) .. (243.82, 638.037);
  \draw[ipe dash dash dotted]
    (150.397, 758.096)
     .. controls (182.75, 601.557) and (230.309, 603.449) .. (244.09, 629.66);
  \draw[ipe dash dash dotted]
    (76.9397, 681.1775)
     .. controls (84.2356, 656.5875) and (96.8009, 653.2095) .. (113.0141, 664.2885)
     .. controls (129.2273, 675.3675) and (149.0883, 700.9035) .. (174.7593, 753.8665);
  \draw[ipe dash dashed]
    (155.8199, 672.066)
     -- (70.6283, 672.066);
  \draw[ipe dash dashed]
    (96.3715, 658.0146)
     -- (156.3604, 658.0146);
  \draw[ipe dash dashed]
    (164.4667, 615.8602)
     -- (224.4552, 615.8602);
  \draw[-{>[ipe arrow small]}]
    (166, 605)
     -- (166, 615);
  \draw[-{>[ipe arrow small]}]
    (156, 648)
     -- (156, 658);
  \draw[-{>[ipe arrow small]}]
    (156, 682)
     -- (156, 672);
  \draw[-{>[ipe arrow small]}]
    (166, 635)
     -- (166, 625);
  \node[ipe node]
     at (137.445, 661.798) {$-e_0 \delta \phi_0$};
  \node[ipe node]
     at (129.311, 614.239) {$-e_0 \delta \phi' $};
\end{tikzpicture}
    \caption{Free energy $G$ along the reaction coordinate of the water reduction sub-step \eqref{eqn:sub1}. The dash-dotted lines show how the free energy curves change due to a change in the potential at the electrode $\delta \phi_0$. $G_\ddag$ is the free energy of the transition state; $G_\mathrm{i}$ and $G_\mathrm{f}$ are the free energies of the initial and final states, respectively. $\Delta_\ddag G$ is the activation energy in the cathodic (reduction) direction. $\Delta_\mathrm{r} G$ is the free energy of the reaction step.}
    \label{fig:reaction-pe-her}
\end{figure}

We describe the initial and final states of the reaction with parabolic potential energy surfaces as a function of some reaction coordinate, as is common in classical transition state theory \citep{Schmickler2010, BardFaulkner2001} -- see Figure \ref{fig:reaction-pe-her}. Our aim is now to express the activation energy $\Delta_\ddag G$ in terms of the applied electrode potential and quantities from the double-layer model. 

To start, the free energies of the equilibrium initial and final state may be written in terms of the electrochemical potentials of the species participating in the reaction \citep{delahay1965}. In our case, we consider reaction sub-step \eqref{eqn:sub1}:
\begin{align}
\begin{split}
    G_\mathrm{i} &= \mu_\mathrm{H_2O} + \mu_\mathrm{e} - e_0 \phi_0 \\
    G_\mathrm{f} &= \mu_\mathrm{H^*} + \mu_\mathrm{OH^-} - e_0 \phi'
\end{split}
\end{align}
where $\phi_0$ is the electric potential at the electrode and $\phi'$ is the electric potential at the reaction plane in the electrolyte. In the common Butler-Volmer model of electrode kinetics \citep{BardFaulkner2001}, the electric potential in the electrolyte is ignored, corresponding to the assumption that $\phi'=0$. Accounting for this potential is then known as applying a Frumkin correction to Butler-Volmer kinetics \citep{frumkin1933wasserstoffuberspannung}.

From the expressions of $G_\mathrm{i,f}$ it follows that the reaction energy of the water reduction sub-step \eqref{eqn:sub1} is
\begin{align} \label{eq:delta-r-g-chempot}
\begin{split}
    \Delta_\mathrm{r} G &= G_\mathrm{f} - G_\mathrm{i} \\
    &= \mu_\mathrm{H^*} + \mu_\mathrm{OH^-} - \mu_\mathrm{H_2O} - \mu_\mathrm{e} + e_0(\phi_0 - \phi').
\end{split}
\end{align}
We will assume that the chemical potentials are independent of the applied electrode potential. Because the chemical potential of a species depends on its concentration \citep{kondepudi1998modern}, this assumption implies that we assume the concentrations of \ce{H^*}, \ce{OH^-} and \ce{H_2O} to be independent of the electrode potential. For hydrogen evolution with a rate-determining Volmer step as considered here, the coverage of \ce{H^*} is likely to be small regardless of the applied potential. On the other hand, \ce{OH^-} is repelled more strongly from the electrode at potentials increasingly negative to the pzc, and the concentration of \ce{H_2O} could be affected by the number of cations accumulating in the double layer. For simplicity, we will neglect these dependencies here, and focus on the effect of the electric potential on the reaction alone. We return to the effect of this simplification when we discuss the results of our model (Sec. \ref{sec:double-layer-structure}).

A variation in the applied potential of $\delta \phi_0$ then only results in a corresponding change in the potential at the reaction plane of $\delta \phi'$. The reaction energy changes accordingly by
\begin{equation}
    \delta(\Delta_\mathrm{r} G) = e_0 \delta (\phi_0 - \phi').
\end{equation}
We can now invoke the Bell-Evans-Polanyi (BEP) principle (\ref{sec:bep}) to find the corresponding change in activation energy;
\begin{equation}
    \delta(\Delta_\ddag G) = \alpha e_0 \delta (\phi_0 - \phi').
\end{equation}
The proportionality factor $\alpha$ can be identified with the transfer coefficient in electrochemical kinetics. Commonly, $\alpha\approx\frac12$ \citep{BardFaulkner2001}.
We assume for simplicity that $\alpha$ is independent of potential. With respect to the pzc (where $\phi_0=\phi'=0$), the activation energy can then be written as
\begin{align} \label{eq:fbv-activation}
\begin{split}
    \Delta_\ddag G &= (\Delta_\ddag G)_\mathrm{pzc} + \alpha e_0(\phi_0 - \phi'), \\
\end{split}
\end{align}
where $(\Delta_\ddag G)_\mathrm{pzc}$ is the activation energy barrier at the pzc. Such a form for the activation energy is usually found in Frumkin-corrected Butler-Volmer theory \citep{delahay1965}. 

With the above expression for the activation energy, we express the cathodic water-reduction-limited hydrogen evolution current density as 
\begin{equation} \label{eq:current-delta-act-g}
    j = -\frac{2 e_0 \bar n_\mathrm{H_2O}}{\beta h} \exp(-\beta \Delta_\ddag G).
\end{equation}
The prefactor is not qualitatively relevant but is motivated as follows. The reaction rate should be proportional to the surface number density of water molecules, $\bar n_\mathrm{H_2O}\approx\qty{5}{nm^{-2}}$ \citep{spaight2020modeling}. The $1/\beta h$ comes from absolute rate theory \citep{BardFaulkner2001} -- $h$ is Planck's constant. The factor 2 is due to the transfer of two electrons in the hydrogen evolution reaction, assuming a Volmer-Heyrovsky mechanism \citep{qin2023cation}. Again, with a mechanism where the Volmer step is rate-determining, the \ce{H^*} coverage is likely low and we can neglect the dependence on the availability of free adsorption sites.

We further assume that the cathodic current dominates the total current and thus neglect any anodic contribution. This assumption is reasonable for gold in particular, on which hydrogen oxidation is highly unfavorable \citep{strmcnik2013improving}. However, even in the experiments on platinum electrodes that we will compare our model to, hydrogen oxidation is suppressed, as the evolved hydrogen gas quickly escapes from the electrolyte. With the activation energy of Eq. \eqref{eq:fbv-activation}, we can thus write the current density as 
\begin{equation} \label{eq:her-fbv}
    j = - \frac{2 e_0 \bar n_\mathrm{H_2O}}{\beta h} \exp \bigg(-\beta (\Delta_\ddag G)_\mathrm{pzc}
    -\alpha \beta e_0 (\phi_0 - \phi') \bigg).
\end{equation}
where $(\Delta_\ddag G)_\mathrm{pzc}$ and $\alpha$ are free parameters. 

In conclusion, we derived the current density by considering how the activation energy depends on the difference in electrostatic potential energy between initial and final states. The interfacial electric field may additionally alter the electronic structure of reactants, which may in turn affect the reaction rate. This effect is neglected here.

Finally, an important difference between our approach and that of \citet{koper2023theory} is that \citeauthor{koper2023theory} interprets the effect of the cation more locally, rather than a mean-field potential at the reaction plane. \citet{qin2023cation} show that the Volmer step energy barrier is not affected by whether a cation directly coordinates the reaction intermediate or not -- contrary to results for carbon dioxide reduction \citep{monteiro2021absence}. This finding suggests that the electrolyte effects on hydrogen evolution may be well-described by variations in the mean-field interfacial electric potential, which is what we consider in this work.

\subsubsection{Hydroxide ion transfer through the double layer} \label{sec:oh-transfer}
Upon the completion of the water reduction sub-step \eqref{eqn:sub1}, we are left with an \ce{OH^-} ion in the reaction plane. The transfer of \ce{OH^-} through the double layer may occur by `bond flipping' or `proton shuttling' in a network of hydrogen-bonded water molecules \citep{tuckerman2002nature, li2022hydrogen}. However, this water structure might be destroyed by strong electric fields \citep{ledezma2017interfacial, sarabia2018effect} or the presence of weakly hydrated cations \citep{botaohuang2021cation, marcus2010effect, li2022hydrogen}, making the transfer process less efficient. If the water structure is altered by such factors, it needs to be reorganized \citep{botaohuang2021cation} in order to transfer the \ce{OH^-} ion from the reaction plane to the electroneutral electrolyte region. 

We propose a simple model, in which we regard the \ce{OH^-} transfer sub-step as another activated process. We stress that this step entails \ce{OH^-} ion transfer through the highly concentrated and charged double layer, not in the diffusion layer -- the effect of the transport of \ce{OH^-} in the diffusion layer has been described by \citet{goyal2021understanding}. We will apply the BEP principle to derive an expression for the rate of \ce{OH^-} transfer through the double layer. The initial state of this transfer step is \ce{OH^-} at the electrode surface (i.e., at $x'$), and the final state is \ce{OH^-} in the electroneutral bulk, where $\phi=0$. The reaction energy of the \ce{OH^-} transfer step is therefore
\begin{align}
\begin{split}
    \Delta_\mathrm{r} G &= G_\mathrm{f} - G_\mathrm{i} \\
    &= \mu_{\mathrm{OH^-}(x=x')} - \mu_{\mathrm{OH^-(bulk)}} + e_0\phi' 
\end{split}
\end{align}
We neglect any potential-dependence in the difference between the chemical potential of $\mu_\mathrm{OH^-}$ at $x=x'$ and in the bulk. According to the BEP principle, the activation energy is then
\begin{equation} \label{eq:pt-activation}
    \Delta_\ddag G = (\Delta_\ddag G)_\mathrm{pzc} + \tilde \alpha e_0 \phi'
\end{equation}
and so the current density can be written
\begin{equation} \label{eq:pt-current}
    j = -A e^{- \tilde \alpha \beta e_0 \phi'}.
\end{equation}
Here, $A$ is some potential-independent prefactor, and $\tilde \alpha$ is a kind of transfer coefficient for the transfer of \ce{OH^-}. 

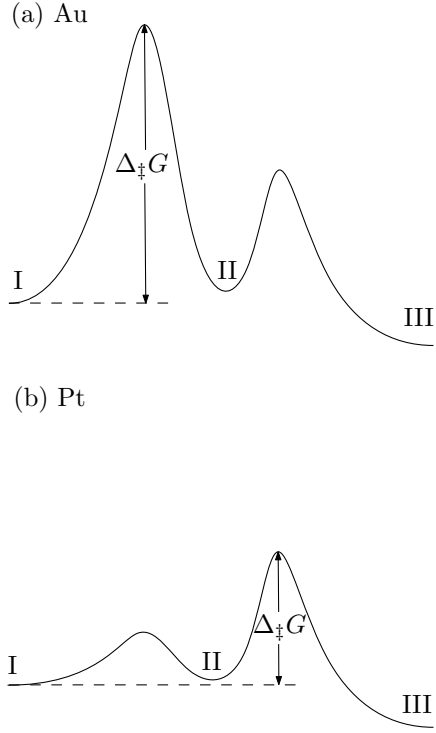
\begin{figure}[t]
    \centering
    \begin{tikzpicture}[ipe stylesheet]
  \draw
    (32, 703.87)
     .. controls (65.5486, 703.87) and (73.9358, 791.935) .. (80.9251, 806.6125)
     .. controls (87.9144, 821.29) and (93.5058, 762.58) .. (100.4952, 733.225)
     .. controls (107.4845, 703.87) and (115.8718, 703.87) .. (121.4632, 716.4507)
     .. controls (127.0547, 729.0313) and (129.8503, 754.1927) .. (134.0438, 754.1927)
     .. controls (138.2373, 754.1927) and (143.8287, 729.0313) .. (153.2693, 712.4832)
     .. controls (162.71, 695.935) and (176, 688) .. (192, 688);
  \draw
    (31.487, 559.87)
     .. controls (65.035, 559.87) and (73.4225, 576.644) .. (80.4119, 579.4397)
     .. controls (87.4013, 582.2353) and (92.9927, 571.0527) .. (99.982, 565.4613)
     .. controls (106.9713, 559.87) and (115.3587, 559.87) .. (120.9502, 572.4507)
     .. controls (126.5417, 585.0313) and (129.3373, 610.1927) .. (133.5308, 610.1927)
     .. controls (137.7243, 610.1927) and (143.3157, 585.0313) .. (152.8846, 568.4832)
     .. controls (162.4535, 551.935) and (176, 544) .. (192, 544);
  \draw[{<[ipe arrow tiny]}-]
    (83.0806, 809.007)
     -- (83.5, 761.754);
  \draw[{<[ipe arrow tiny]}-]
    (133.471, 610.159)
     -- (133.656, 587.498);
  \node[ipe node, font=\footnotesize]
     at (71.861, 753.452) {$\Delta_\ddag G$};
  \node[ipe node, font=\footnotesize]
     at (123.87, 579.727) {$\Delta_\ddag G $};
  \draw[-{>[ipe arrow tiny]}]
    (83.1981, 749.153)
     -- (83.5905, 703.962);
  \draw[-{>[ipe arrow tiny]}]
    (133.656, 575.377)
     -- (133.729, 560.116);
  \node[ipe node, font=\footnotesize]
     at (33.729, 709.457) {I};
  \node[ipe node, font=\footnotesize]
     at (110.61, 712.967) {II};
  \node[ipe node, font=\footnotesize]
     at (180.821, 695.415) {III};
  \node[ipe node, font=\footnotesize]
     at (31.649, 564.053) {I};
  \node[ipe node, font=\footnotesize]
     at (104.669, 565.457) {II};
  \node[ipe node, font=\footnotesize]
     at (180.146, 549.309) {III };
  \node[ipe node, font=\footnotesize]
     at (32.726, 810.138) {(a) Au
};
  \node[ipe node, font=\footnotesize]
     at (33.776, 665.246) {(b) Pt

};
  \draw[ipe dash dashed]
    (32, 704)
     -- (96, 704);
  \draw[ipe dash dashed]
    (32, 560)
     -- (144, 560);
\end{tikzpicture}
    \caption{Proposed energy landscape over the course of the reaction for the Volmer step on (a) gold and (b) platinum. The numbers I-III denote intermediate states throughout the reaction, defined as follows. I: \ce{H_2O + $e^-$}; II: \ce{OH^-($x=x'$) + H^*}, III: \ce{OH^-(bulk) + H^*}. Hence, the first barrier is for water reduction, and the second barrier for \ce{OH^-} transfer.}
    \label{fig:energy-diagrams}
\end{figure}

\subsection{Alkaline hydrogen evolution on gold and platinum} \label{sec:gold-and-platinum}
To explain the different trends for the alkaline hydrogen evolution reaction observed on gold and platinum electrodes, we propose the energy landscape shown in Fig. \ref{fig:energy-diagrams}, which builds on the ideas of \citet{bender2022understanding}. For gold, the transfer of \ce{OH^-} presents a barrier that is rather small as compared to water reduction and does not affect the overall activation energy. 

We interpret the high current density on platinum electrodes measured in experiments as being due to a much lower water reduction energy barrier.\footnote{The situation is complicated by the existence of multiple adsorbed hydrogen species on Pt -- see \citet{Schmickler2010}, Ch. 14 for a discussion.} For platinum, we therefore consider that the activation energy is dominated by the \ce{OH^-} transfer energy barrier. Note that the water reduction barrier depends on the electric potential in the double layer through Eq. \eqref{eq:fbv-activation} and the \ce{OH^-} transfer barrier depends on the electric potential through Eq. \eqref{eq:pt-activation}.

\section{Results and discussion}
In the following, we first discuss the behavior of the electric potential and ion concentrations in the double layer as computed by solving Eq. \eqref{eq:pbe} with Eq. \eqref{eq:stern-bc} as boundary condition. We then compute the current density using the relevant values for the electric potential, and compare the results to experimental data.

\subsection{Double-layer structure} \label{sec:double-layer-structure}
\begin{figure*}[t]
    \centering
    \includegraphics[width=\textwidth]{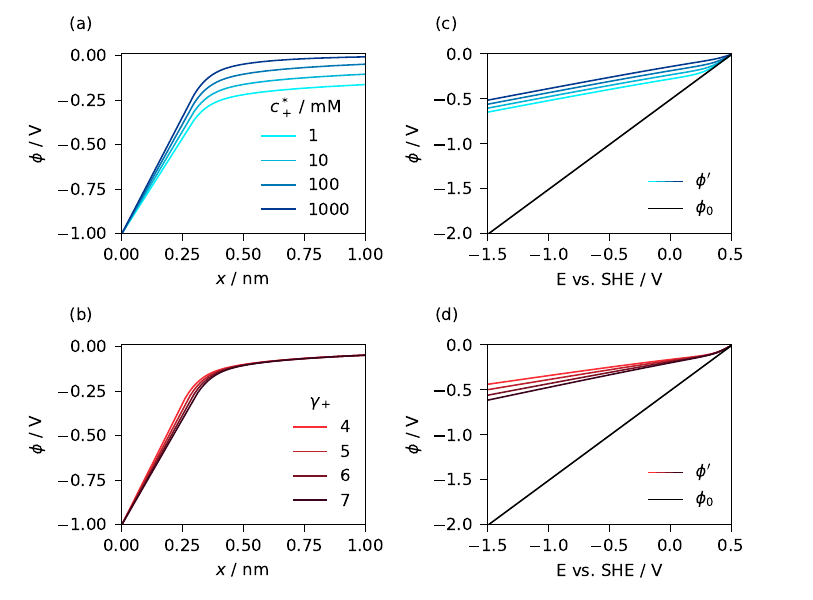}
    \caption{(a-b) Potential profiles at $\phi_0=\qty{-1}{V}$ calculated from the double-layer model for varying cation bulk concentration $c_+^*$ (a) and effective cation size $\gamma_+$ (b).
    (c-d) Potential at the reaction plane $\phi'$ calculated for gold electrodes for various $c_+^*$ (c) and for various $\gamma_+$ (d), plotted against the electrode potential on the SHE scale. The potential at the electrode $\phi_0$ is shown with a black line. 
    Unless otherwise specified in the legends, $\gamma_+=6$ and $c_+^*=\qty{100}{mM}$. In figures (c-d), the value for the pzc used is for gold electrodes ($\mathsf E_\mathrm{pzc,Au}=\qty{0.51}{V}$ vs. SHE). }
    \label{fig:phi0-phiprime}
\end{figure*}

Central to our approach is the behavior of the electric potential in the double layer. Figure \ref{fig:phi0-phiprime}a and b show the electric potential profiles obtained from the double-layer model for various cation bulk concentrations $c_+^*$ and effective cation sizes $\gamma_+$ at $\phi_0=\qty{-1}{V}$. We observe that the potential decays faster for higher cation bulk concentrations, and smaller effective cation sizes. Smaller cations pack more tightly in the double layer, shielding the surface charge more effectively. 

From such potential profiles, we can read off the potential at the reaction plane $\phi'$ at $x'\approx\qty{0.28}{nm}$, see Sec. \ref{sec:electron-transfer}. Let us consider how $\phi'$ behaves with the electrode potential $\mathsf{E}$. From $\mathsf{E}$ we find $\phi_0$ by Eq. \eqref{eq:pzc-phi0}, which defines our boundary condition. Figure \ref{fig:phi0-phiprime}c and d show the behavior of $\phi'$ against $\mathsf{E}$ on gold electrodes for various $c_+^*$ and $\gamma_+$. We see that smaller cation bulk concentrations and larger cations yield a larger $\phi'$. As a result, the potential difference $\phi_0 - \phi'$ increases with increasing cation bulk concentration $c_+^*$ and decreasing cation size $\gamma_+$. 

A different pzc would only horizontally shift the curves in Fig. \ref{fig:phi0-phiprime}c and d. As discussed in Sec. \ref{sec:potentials}, hydrogen always evolves negative of the pzc. In this potential range, the trends in $\phi'$ with electrolyte composition are thus the same for both gold and platinum. In our model, a change in pzc can therefore not explain the reversed cation trends on gold and platinum. For this reason, it is necessary to consider the change in rate-determining step proposed in Sec. \ref{sec:gold-and-platinum}.

\begin{figure}
    \centering
    \includegraphics[width=\textwidth]{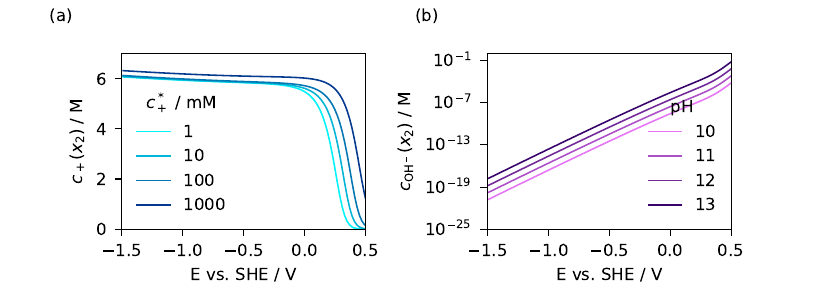}
    \caption{(a) Cation concentration at the plane of closest approach $c_+(x_2)$ against the electrode potential, calculated for various $c_+^*$. (b) Hydroxide ion concentration at $x_2$ against the electrode potential, computed for various values of the bulk pH.
    Parameters: $\mathsf E_\mathrm{pzc,Au}=\qty{0.51}{V}$ vs. SHE, $\gamma_+=6$, and unless otherwise specified in the legend, $c_+^*=\qty{100}{mM}$. }
    \label{fig:concentration-epot}
\end{figure}

Next, we study the interfacial cation concentration. Figure \ref{fig:concentration-epot}a shows that the interfacial cation concentration near the electrode surface takes an almost constant value at electrode potentials far from the pzc, regardless of the bulk concentration. Various studies \citep{goyal2021interrelated, monteiro2021understanding, ringe2023cation} presume that an increase in the interfacial cation concentration with increasing bulk concentration and with more negative electrode potentials is responsible for the kinetic cation effects.  Based on the results of this mean-field model, though, it is unlikely that cation effects can be explained by the interfacial cation concentration alone. 

At the pzc ($\mathsf{E} = \qty{0.51}{V}$ vs. SHE in Fig. \ref{fig:concentration-epot}a), the mean-field model predicts that the interfacial cation concentration is equal to the bulk concentration. However, measurements of the differential capacitance suggest that the interfacial ion concentration at the pzc can be on the order of $3$ to \qty{4}{M} at a bulk concentration of \qty{50}{mM} \citep{garlyyev2018influence, xue2020nature}. The discrepancy between the mean-field model considered here and experiments may be explained by weak cation-surface interactions \citep{schmickler2021effect, ojha2022double}. Nevertheless, because we consider potentials much more negative from the pzc, we expect that the effect of these weak interactions is negligible compared to the strong electrostatic interactions.

Finally, the concentration of \ce{OH^-} in the double layer is plotted against $\mathsf{E}$ for various bulk pH values in Figure \ref{fig:concentration-epot}b. At an electrode potential of \qty{-1}{V} vs. SHE, the \ce{OH^-} concentration is around the order of $10^{-15}$ M, meaning that \qty{1}{\mu m^3} of electrolyte contains about one \ce{OH^-} ion. At such homeopathic concentrations, the number of \ce{OH^-} molecules in the double layer (at most a few nm thick) is essentially zero. Hence, our assumption that the concentration of \ce{OH^-} is independent of the electrode potential is reasonable for sufficiently negative potentials. Modeling the concentration of \ce{OH^-} at non-equilibrium conditions is left for future work.

\begin{figure*}[t!]
    \centering
    \includegraphics[width=\textwidth]{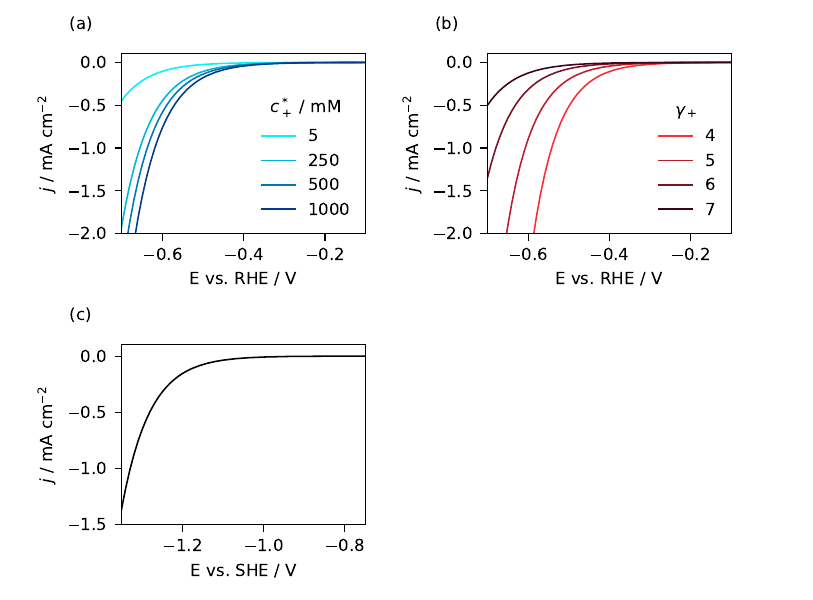}
    \caption{Hydrogen evolution current density on gold, calculated according to Eq. \eqref{eq:her-fbv} where $(\Delta_\ddag G)_\mathrm{pzc}=\qty{1.37}{eV}$ and $\alpha=\frac12$. (a) Current density for various cation bulk concentrations $c_+^*$, shown on the RHE scale for pH 11. (b) Current density for various effective cation sizes, shown on the RHE scale for pH 11. (c) pH-independent current density on the SHE scale. Unless indicated otherwise in the legend, $c_+^*=\qty{100}{mM}$ and $\gamma_+=6$.}
    \label{fig:gold-fbv}

\end{figure*}

\floatsetup[figure]{style=plain,subcapbesideposition=top}
\begin{figure*}
    \centering
    \sidesubfloat[]{
        \label{fig:gold-conc-trend}
        \includegraphics[height=110px]{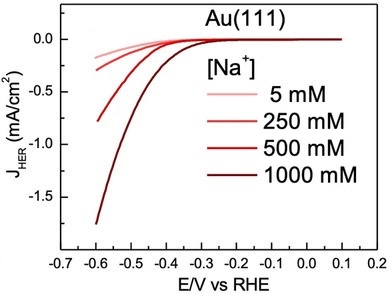}
    } 
    \sidesubfloat[]{
        \label{fig:gold-species-trend}
        \includegraphics[height=110px]{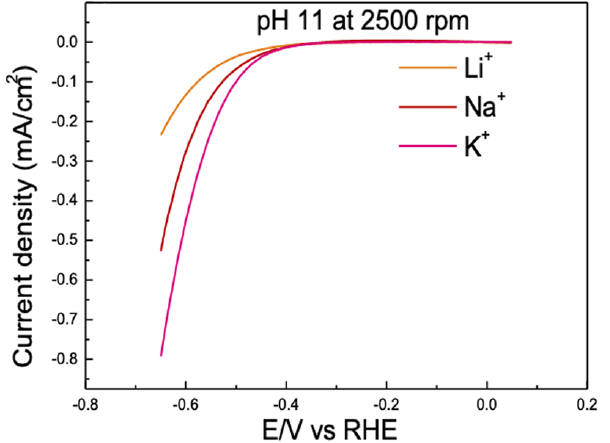}
    } \\
    \sidesubfloat[]{
        \label{fig:gold-ph-trend}
        \includegraphics[height=130px]{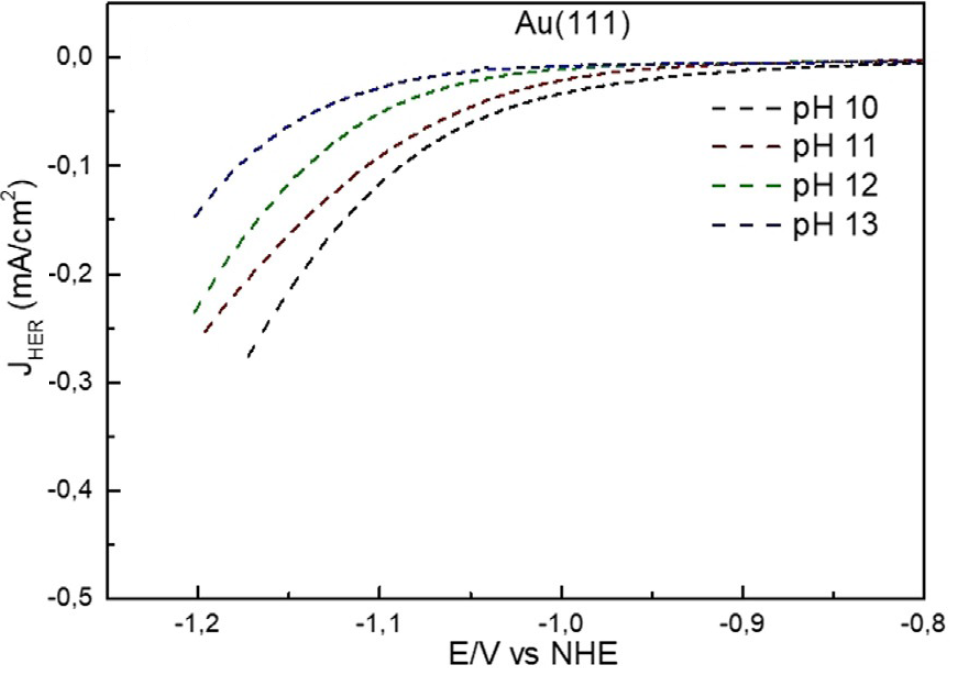}
    } \\
    \caption{Experimentally obtained voltammograms (current density vs. electrode potential) showing various electrolyte effects for the alkaline HER on gold electrodes. For all experiments, the electrode rotation rate is 2500 rpm, and the scan rate \qty{25}{mV/s}. (a) Cation concentration series on \ce{Au(111)} for \ce{NaClO4} in \qty{1}{mM} \ce{NaOH} (pH $11$). (b) Cation species series on a polycrystalline gold electrode with a cation bulk concentration of $c_+^*=\qty{100}{mM}$ at pH 11. (c) pH series on \ce{Au(111)} for $x$ M \ce{NaOH} and $(0.1 - x)$ M \ce{NaClO4}, where $x=0.1, 0.01, 0.001$ and $0.0001$, corresponding to pH 13, 12, 11 and 10 respectively. Note that the electrode potential is on the NHE (Normal Hydrogen Electrode) scale, equivalent to SHE.  Figures (a) and (c) are reprinted from \citet{goyal2021interrelated} under a CC BY 4.0 license. Figure (b) is reprinted from \citet{goyal2021understanding} with permission from AIP Publishing. }
    \label{fig:gold-experimental}
\end{figure*}

\subsection{Trends on gold}
Recall that for hydrogen evolution on gold, we expect that the rate is limited by the water reduction sub-step \eqref{eqn:sub1} (Sec. \ref{sec:gold-and-platinum}). The current density is thus described by Eq. \eqref{eq:her-fbv}. Figure \ref{fig:gold-fbv} shows the current density as computed from Eq. \eqref{eq:her-fbv} and the double-layer model. We took pH $11$, a cation concentration of $c_+^*=\qty{100}{mM}$ and $\gamma_+=6$ (corresponding to \ce{Na^+}), unless otherwise indicated in the legend. We also chose $(\Delta_\ddag G)_\mathrm{pzc}=\qty{1.37}{eV}$ and $\alpha=\frac12$ to obtain realistic current density values. 

As expected from the behavior of $\phi_0 - \phi'$ discussed before, the current density also increases with increasing cation bulk concentration (Fig. \ref{fig:gold-fbv}a) and with decreasing cation effective size (Fig. \ref{fig:gold-fbv}b). These trends are qualitatively the same as those in the experimental data from \citet{goyal2021interrelated, goyal2021understanding}, which is reprinted in Fig. \ref{fig:gold-experimental}a and b, and other experimental work \citep{bender2022understanding, xue2018influence}. Note that the experimental conditions match the parameters in the model calculations. An increase in current with decreasing cation effective size was also reported by \citet{ringe2019understanding} for carbon dioxide reduction, based on a double-layer model similar to the one used here. 

Interestingly, for a single choice of $(\Delta_\ddag G)_\mathrm{pzc}$ and $\alpha$, the calculated current density values match the data rather well, even across different experiments. Nevertheless, there are some quantitative differences. Consider the cation bulk concentration series, for example. Between $c_+^*=\qty{5}{mM}$ and \qty{250}{mM}, the simulated current density increases significantly, whereas the measured current density does not increase as much. On the other hand, increasing $c_+^*$ from \qty{250}{mM} to \qty{1000}{mM} leads to a small increase in the simulations, whereas a big increase is measured experimentally. We will analyze the quantitative performance of the model in more detail in Sec. \ref{sec:quantitative}.

The current according to Eq. \eqref{eq:her-fbv} is independent of the bulk pH. Figure \ref{fig:gold-ph-trend} shows the current on the SHE scale for $\gamma_+=6$ and $c_+^*=\qty{100}{mM}$. The predicted pH-independence agrees with experimental data of \citet{strmcnik2013improving}. However, \citet{goyal2021interrelated} do measure a pH effect on gold electrodes, as shown by the reprinted plot in Fig. \ref{fig:gold-ph-trend}. In this data, the onset of hydrogen evolution shifts to more negative potentials for higher pH. This may be due to a bulk pH-dependence of the (non-equilibrium) concentration of interfacial \ce{OH^-}.

It is also possible to express the current density using Marcus theory. The derivation and the corresponding results are shown in \ref{app:marcus}. The qualitative behavior is the same as in Fig. \ref{fig:gold-fbv}, and the above discussion applies to these results as well.

\subsection{Trends on platinum} \label{sec:platinum}
\begin{figure}[ht!]
    \centering
    \includegraphics[width=\linewidth]{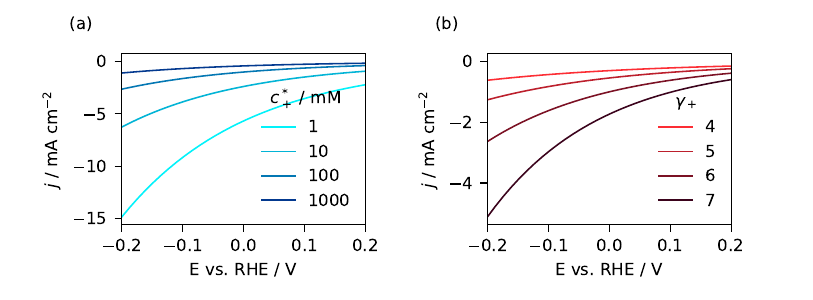}
    \caption{Calculated hydrogen evolution current density with \ce{OH^-} transfer as rate-determining step. (a) Current density against electrode potential on RHE scale for various cation bulk concentrations $c_+^*$. (b) Current density for various effective ion sizes $\gamma_+$. Unless otherwise indicated in the legend, $c_+^*=\qty{100}{mM}$ and $\gamma_+=6$. Parameters: $A_\mathrm{Pt}=\frac{2e_0 \bar n_\mathrm{H_2O}}{\beta h} \exp(-\beta C)$, with $C=\qty{0.87}{eV}$ -- note that this is merely a convenient form with little physical meaning. The potential range was chosen similarly to the Pt data in Fig. \ref{fig:bender-au-pt}.}
    \label{fig:pt-current}
\end{figure}

Next, we calculate the current density on platinum electrodes, where we expect that \ce{OH^-} transfer through the double layer limits the rate. Hence, we use Eq. \eqref{eq:pt-current}, and $\mathsf{E}_\mathrm{pzc}$ for platinum. Figure \ref{fig:pt-current} shows the result for various cation bulk concentrations, cation species, and pH. Fig. \ref{fig:pt-current}a and b show that the cation species and concentration trends are reversed compared to gold.  

\begin{figure}[ht!]
    \centering
    \includegraphics[height=120px]{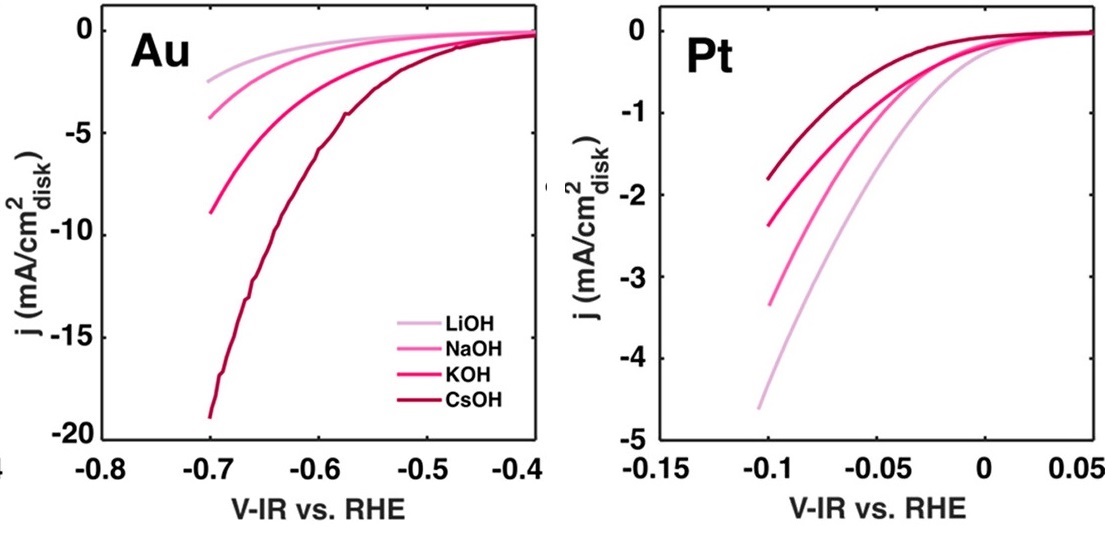}
    \caption{Experimental voltammograms showing the cation species trends on polycrystalline gold (Au) and polycrystalline platinum (Pt) in \qty{0.1}{M} electrolytes (pH 13). Reprinted with permission from \citet{bender2022understanding}. Copyright 2022 American Chemical Society.}
    \label{fig:bender-au-pt}
\end{figure}

The reversal of the cation species trend on platinum is consistent with experimental data of \citet{bender2022understanding}, which is reprinted in Figure \ref{fig:bender-au-pt}. Furthermore, \citet{monteiro2021understanding} showed that the current density decreases for larger cation concentrations on platinum at pH $11$ and higher. This observation agrees with the reversed cation concentration trend predicted by the model introduced here. Hence, although the model is rather simple -- we only considered \ce{OH^-} transfer as an activated process -- it qualitatively reproduces the experimentally observed reversal of cation trends. 

\ce{OH^-} transfer can only be rate-limiting if enough \ce{OH^-} is produced, so water reduction would in reality still limit the rate at low current densities. It is interesting to note that \citet{monteiro2021understanding} also report a reversal of the cation species trend on gold electrodes at large negative overpotentials. In the diagram of Figure \ref{fig:energy-diagrams}, such a trend reversal can be expected on gold electrodes if the water reduction barrier is lowered so far (by a large value of $\phi_0 - \phi'$) that the \ce{OH^-} transfer barrier dominates. 

An important step towards developing a more quantitatively accurate model will be to include some notion of the water structure, and how it is affected by various cations \citep{botaohuang2021cation, li2022hydrogen, briega2021cation} and the electric field \citep{ledezma2017interfacial, sarabia2018effect}. Measurements of \citet{briega2021cation} suggest that the pzfc may shift to slightly more positive values for more weakly hydrated cations, which would bring the curves in Fig. \ref{fig:pt-current}b more closely together. In addition, the rate of \ce{OH^-} transfer in the double layer rate likely depends on the bulk pH \citep{goyal2021understanding}, which may explain the observed pH-dependence \citep{strmcnik2013improving} on platinum electrodes. Finally, it remains unclear how the accumulation of \ce{OH^-} in the double layer can limit the reaction rate -- the effect of interfacial \ce{OH^-} on the water reduction reaction energetics needs to be clarified.

\subsection{Quantitative comparison} \label{sec:quantitative}
Lastly, we consider the relationship between $j$ and $\phi_0 - \phi'$ in more detail. Taking the base-10 logarithm of Eq. \eqref{eq:her-fbv}, we obtain a Tafel-like relation
\begin{equation} \label{eq:tafel-like}
    \log |j| = a - b (\phi_0 - \phi').
\end{equation}
Here, the parameters $a$ and $b$ contain all potential-independent parameters, most of which we do not know (e.g., the exponential prefactor). $b$ is related to the transfer coefficient $\alpha$ as 
\begin{equation}
    b = \frac{\alpha \beta e_0}{\ln 10}.
\end{equation}

To further evaluate the validity of our model, we look for a relationship like Eq. \eqref{eq:tafel-like} in experimental data. We use the data of \citet{goyal2021understanding}, who measured the hydrogen evolution current on gold for different cation concentrations, cation species, and pH at various electrode potentials. For each combination of experimental conditions, we compute $\phi_0 - \phi'$ from the double-layer model. Based on Eq. \eqref{eq:tafel-like} we then expect a linear relation between the measured current density $j$ and the calculated $\phi_0 - \phi'$. Data for different electrolytes should fall on the same line, assuming that $a$ does not depend on the electrolyte composition.

\begin{figure}
    \centering
    \includegraphics[width=\linewidth]{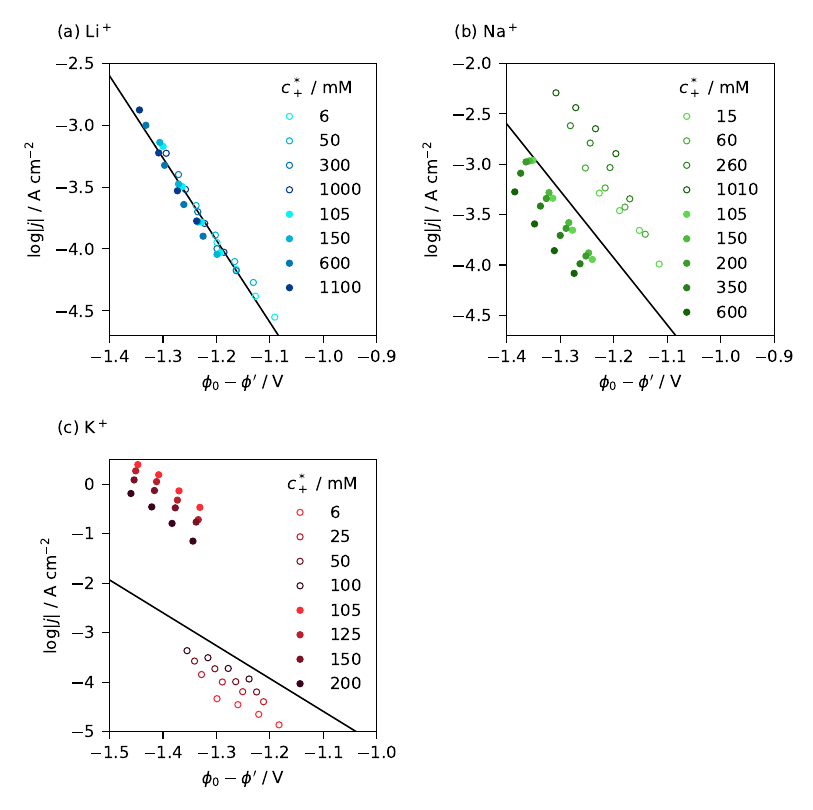}
    \caption{The logarithm of the measured current density on polycrystalline gold electrodes, $\log |j|$, against the potential drop in the double layer, $\phi_0 - \phi'$, computed from the double-layer model for the corresponding experimental conditions. Data obtained from \citet{goyal2021understanding}. Subfigure labels indicate the cation species. Open symbols: pH 11, filled symbols: pH 13. Parameters: $\gamma_+=7$ for \ce{Li^+}, $6$ for \ce{Na^+}, $5$ for \ce{K^+}. The fit of the \ce{Li^+} data in subfigure (a) with $\alpha=0.39$ is shown as a black line in all subfigures. }
    \label{fig:quantitative}
\end{figure}

In Figure \ref{fig:quantitative}, the current density measured by \citeauthor{goyal2021understanding} is plotted against $\phi_0 - \phi'$ as calculated from our mean-field model for the different experimental conditions. The subfigures a, b, and c show data for \ce{Li^+}-, \ce{Na^+}- and \ce{K^+}-based electrolytes, each for various concentrations. For each concentration, a series of data points at four different electrode potentials is shown. Data for pH 11 is shown with open symbols, whereas data for pH 13 is represented with filled symbols.

The data for \ce{Li^+} electrolytes in Fig. \ref{fig:quantitative}a falls on a straight line, demonstrating the validity of Eq. \eqref{eq:her-fbv}. From a linear fit we obtain the transfer coefficient $\alpha=0.39$. The linear fit is indicated with a black line, and it is reproduced in subfigures b and c as well. In contrast to the \ce{Li^+} data, the data points for \ce{Na^+} and \ce{K^+} do not collapse on one line. Moreover, for \ce{K^+} the pH 13 data shows a much larger current than the pH 11 data, whereas the \ce{Na^+} data shows a slightly higher current for the pH 11 data. 

To explain why the model only describes the data for \ce{Li^+}, we note that \ce{Li^+} is a `water-structure-making' ion, whereas \ce{Na^+} and especially \ce{K^+} break the water structure \citep{marcus2010effect}. A broken water structure would hamper \ce{OH^-} transfer. We suggest that our model for the water-reduction-limited current is accurate as long as the water structure is maintained (so that \ce{OH^-} transfer is not hampered), similar to what has been suggested by \citet{ringe2023cation}. Another consideration is that the smaller solvation shell of \ce{Na^+} and \ce{K^+} allows for more short-ranged interactions between cation and reactant, as considered by \citet{koper2023theory}. The large solvation shell of \ce{Li^+} may only allow long-ranged electrostatic interactions, which are the focus of our model. 

The current measurements on platinum electrodes from \citet{monteiro2021understanding} can be analyzed in a similar way (see \ref{app:platinum}). The data for \ce{Li^+} at pH 13 again collapses on one line (with $\alpha=0.36)$, whereas the data for pH 11 shows a slight deviation. The \ce{K^+} data for pH 9 with low cation concentrations seems to fit the trend of \ce{Li^+} at pH 13. Above pH 10, increasing the cation concentration has a strong inhibitory effect. To describe this data, one would need to model the local \ce{OH^-} concentration in nonequilibrium conditions and its effect on the reaction rate. It should also be noted that recent work of \citet{goyal2024cooperative} suggests that cations only affect hydrogen evolution on Pt electrodes at stepped sites. On Pt(111), having no stepped sites, \citeauthor{goyal2024cooperative} measure no cation effect at all, in stark contrast to the data on Au(111) shown in Fig. \ref{fig:gold-experimental}. These findings emphasize that the double-layer structure of platinum electrodes and its effect on kinetics requires further study.

The described behavior is qualitatively similar for different choices of parameters. Because \citet{ringe2019understanding} use rather different values for $x_2$, we show the results for our model with their values of $x_2$ in \ref{app:ringe}. This parameter choice brings the \ce{K^+} data at pH 13 more in line with the \ce{Li^+} data. However, since the accuracy of the model does not improve for the other data, the evidence is insufficient to conclude that these parameters are more realistic.

\section{Conclusion and outlook}
In this work, we developed expressions for the alkaline hydrogen evolution current density in terms of the electric potential in the double layer, which was computed using a mean-field double-layer model. We assume that the Volmer step can be split into two sub-steps, the water reduction sub-step \eqref{eqn:sub1} and the \ce{OH^-} transfer sub-step \eqref{eqn:sub2}. The model for the water-reduction-limited sub-step explains the increase of the current with the cation bulk concentration and more weakly hydrated cation species. It also correctly predicts the pH-independence of the water-reduction-limited current. With the simple model we proposed for the \ce{OH^-}-transfer-limited current, we can reproduce the experimentally observed cation trend reversal between gold and platinum electrodes. 

Moreover, the model for the water-reduction-limited current describes the current in \ce{Li^+}-based electrolytes with the correct functional form. In agreement with previous research, we propose that this model is valid when the water structure is sufficiently maintained to support the transfer of \ce{OH^-}. When the water structure is destroyed, the transfer of \ce{OH^-} limits the rate. 

The approach taken here provides simple analytic expressions and an intuitive physical view. In addition, the theory can serve as a reference point for deeper investigations. Further research should focus on the inhibitory effect of cations on the reaction rate, for example through their effect on \ce{OH^-} transfer through the interfacial water structure. The understanding of the double layer at platinum electrodes needs to be refined as well, e.g., the approach of cations to the surface, their interaction with adsorbates, and the surface charge distribution around steps.

\section*{CRediT author statement}
\textbf{Lucas B.T. de Kam}: Methodology, Software, Formal analysis, Investigation, Writing - Original Draft, Visualization; \textbf{Thomas L. Maier}: Conceptualization, Supervision, Writing - Review \& Editing; \textbf{Katharina Krischer}: Conceptualization, Supervision, Writing - Review \& Editing

\section*{Acknowledgements}
The authors kindly thank Dr. Jun Huang (FZ Jülich, Germany) for his help with the implementation of the double-layer model, and Prof. Dr. Christopher Stein (Department of Chemistry, TUM, Germany) for valuable discussions. The authors gratefully acknowledge the support by the Deutsche Forschungsgemeinschaft (DFG, German Research Foundation) through `e-conversion Cluster of Excellence’, EXC 2089/1-390776260, and the support by the Bavarian State Ministry of Science and the Arts within the Collaborative Research Network `Solar Technologies go Hybrid (SolTech)’.

\appendix
\renewcommand{\thefigure}{S\arabic{figure}}
\renewcommand{\theequation}{S\arabic{equation}}
\renewcommand{\thetable}{S\arabic{table}}
\renewcommand{\thesection}{SI \arabic{section}}
\setcounter{figure}{0}

\section{Bell-Evans-Polanyi principle}\label{sec:bep}
\begin{figure}[h!]
    \centering
    \begin{tikzpicture}[ipe stylesheet]
  \draw[shift={(79.706, 731.023)}, xscale=0.394, yscale=0.3907, ipe dash dashed]
    (0, 0) rectangle (0, 0);
  \draw[shift={(22.969, 668.506)}, xscale=1.0943, yscale=0.9302, ->]
    (0, 0)
     -- (150.8177, 0.004);
  \draw[shift={(22.969, 668.506)}, xscale=1.0943, yscale=0.9302, ->]
    (0, 0)
     -- (0, 161.312);
  \node[ipe node, font=\footnotesize]
     at (105.127, 659.887) {reaction coordinate
};
  \node[ipe node]
     at (7.996, 815.738) {$G 
$};
  \node[ipe node, font=\footnotesize]
     at (42.607, 766.819) {$\Delta_\ddag G _\mathrm{f}$
};
  \draw[shift={(31.813, 706.859)}, xscale=0.9188, yscale=0.781, ipe dash dashed]
    (0, 0)
     -- (150.3385, -0.16);
  \draw[shift={(53.889, 775.761)}, xscale=0.9188, yscale=0.781, -{>[ipe arrow tiny]}]
    (0, 0)
     -- (-0.048, 19.022);
  \draw[shift={(53.935, 761.709)}, xscale=1.0927, yscale=1.0936, -{>[ipe arrow tiny]}]
    (0, 0)
     -- (-0.0261, -15.369);
  \draw[shift={(107.795, 793.042)}, xscale=0.9188, yscale=0.781, ipe dash dashed]
    (0, 0)
     -- (-83.1798, 0.122);
  \draw[shift={(109.707, 793.042)}, xscale=0.3938, yscale=0.3347, ipe dash dashed]
    (0, 0)
     -- (195.346, 0);
  \draw[shift={(169.968, 755.806)}, xscale=0.9188, yscale=0.781, -{>[ipe arrow tiny]}]
    (0, 0)
     -- (-0.068, 44.869);
  \draw[shift={(169.952, 742.957)}, xscale=0.9188, yscale=0.781, -{>[ipe arrow tiny]}]
    (0, 0)
     -- (0.041, -43.963);
  \draw[shift={(53.68, 725.909)}, xscale=1.0927, yscale=1.0936, -{>[ipe arrow tiny]}]
    (0, 0)
     -- (0.1024, 7.191);
  \draw[shift={(53.682, 715.681)}, xscale=1.0927, yscale=1.0936, -{>[ipe arrow tiny]}]
    (0, 0)
     -- (0.0355, -6.473);
  \node[ipe node, font=\footnotesize]
     at (158.66, 747.957) {$\Delta_\ddag G_\mathrm{b} $
};
  \node[ipe node, font=\footnotesize]
     at (43.619, 717.882) {$\Delta_\mathrm{r} G $
};
  \draw[shift={(32.041, 760.687)}, xscale=0.9188, yscale=0.781, ipe pen heavier]
    (0, 0)
     .. controls (8.5556, -23.567) and (24.3711, -26.7535) .. (41.1418, -15.7795)
     .. controls (57.9125, -4.8055) and (75.6385, 20.329) .. (101.3095, 73.292);
  \draw[shift={(103.522, 818.561)}, xscale=0.9188, yscale=0.781, ipe pen heavier]
    (0, 0)
     .. controls (30.805, -158.079) and (78.364, -156.187) .. (92.145, -129.976);
  \draw[shift={(32.097, 752.197)}, xscale=0.9188, yscale=0.781, ipe pen heavier, ipe dash dash dotted]
    (0, 0)
     .. controls (8.5556, -23.567) and (24.3711, -26.7535) .. (41.1418, -15.7795)
     .. controls (57.9125, -4.8055) and (75.6385, 20.329) .. (101.3095, 73.292);
  \draw[shift={(106.409, 725.9)}, xscale=0.9188, yscale=0.781, -{>[ipe arrow tiny]}]
    (0, 0)
     -- (0.083, 9.966);
  \node[ipe node, font=\footnotesize]
     at (90.222, 719.862) {$\delta(\Delta_\mathrm{r} G )$
};
  \draw[shift={(106.702, 744.781)}, xscale=0.9188, yscale=0.781, {<[ipe arrow tiny]}-]
    (0, 0)
     -- (0.083, 9.966);
  \draw[shift={(229.031, 737.19)}, xscale=1.0943, yscale=0.9302, ipe pen heavier]
    (0, 0)
     -- (62.76, 85.871);
  \draw[shift={(319.843, 668.037)}, xscale=1.0943, yscale=0.9302, ipe pen heavier]
    (0, 0)
     -- (-59.444, 160.186);
  \draw[shift={(239.986, 667.839)}, cm={0.6615,0,-0.005,1.2834,(0,0)}, ipe pen heavier, ipe dash dash dotted]
    (0, 0)
     -- (193.459, 116.115);
  \draw[shift={(250.884, 784.504)}, cm={0.6615,0,-0.005,1.2834,(0,0)}, ipe dash dashed]
    (0, 0)
     -- (176, 0);
  \draw[shift={(251.204, 702.366)}, cm={0.6615,0,-0.005,1.2834,(0,0)}, ipe dash dashed]
    (0, 0)
     -- (176, 0);
  \draw[shift={(278.385, 729.622)}, cm={0.6615,0,-0.005,1.2834,(0,0)}, ipe dash dashed]
    (0, 0)
     -- (134.841, -0.2);
  \draw[shift={(361.774, 749.348)}, xscale=1.0943, yscale=0.9302, -{>[ipe arrow tiny]}]
    (0, 0)
     -- (0.099, -18.814);
  \node[ipe node, font=\footnotesize]
     at (320.084, 754.453) {$(1-\alpha)
 \delta(\Delta_\mathrm{r}G )$};
  \draw[shift={(269.522, 701.916)}, cm={0.5547,0,-0.005,1.2834,(0,0)}, {<[ipe arrow tiny]}-]
    (0, 0)
     -- (0, 64);
  \node[ipe node, font=\footnotesize]
     at (233.441, 733.588) {$\delta (\Delta_\mathrm{r} G )
$};
  \draw[shift={(361.947, 727.728)}, xscale=1.0943, yscale=0.9302, {<[ipe arrow tiny]}-]
    (0, 0)
     -- (0.046, -6.93);
  \node[ipe node, font=\footnotesize]
     at (335.846, 713.27) {$\alpha  \delta(\Delta_\mathrm{r}G )$};
  \draw[shift={(101.048, 800.426)}, xscale=1.1261, yscale=0.9598]
    (0, 0) rectangle (16.213, -26.482);
  \draw[shift={(361.628, 781.934)}, xscale=1.0943, yscale=0.9302, {<[ipe arrow tiny]}-]
    (0, 0)
     -- (0.133, -19.221);
  \draw[shift={(362.357, 710.067)}, xscale=1.0943, yscale=0.9302, -{>[ipe arrow tiny]}]
    (0, 0)
     -- (0.046, -6.93);
  \node[ipe node, font=\footnotesize]
     at (6.141, 831.26) {(a)};
  \node[ipe node, font=\footnotesize]
     at (211.42, 829.951) {(b)};
  \node[ipe node, font=\footnotesize]
     at (37.357, 688.252) {reactant};
  \node[ipe node, font=\footnotesize]
     at (153.679, 689.089) {product};
  \draw[shift={(31.506, 734.958)}, xscale=1.0943, yscale=0.9302, ipe dash dashed]
    (0, 0)
     -- (75.564, -0.074);
  \draw[shift={(31.598, 743.477)}, xscale=1.0943, yscale=0.9302, ipe dash dashed]
    (0, 0)
     -- (75.564, -0.074);
  \draw[shift={(228.887, 817.243)}, xscale=1.0231, yscale=0.9302]
    (0, 0) rectangle (151.634, -160.402);
\end{tikzpicture}
    \caption{(a) Free energy $G$ over the course of a reaction. $\Delta_\ddag G_\mathrm{f,b}$ are the forward and backward activation energies. $\Delta_\mathrm{r}
    G$ is the reaction energy in the forward direction. When tuning a certain parameter, the free energy surface changes from the solid line to the dash-dotted line, which changes the reaction energy by an amount $\delta(\Delta_\mathrm{r} G)$. (b) Zoom of (a) around the transition state. }
    \label{fig:bep}
\end{figure}
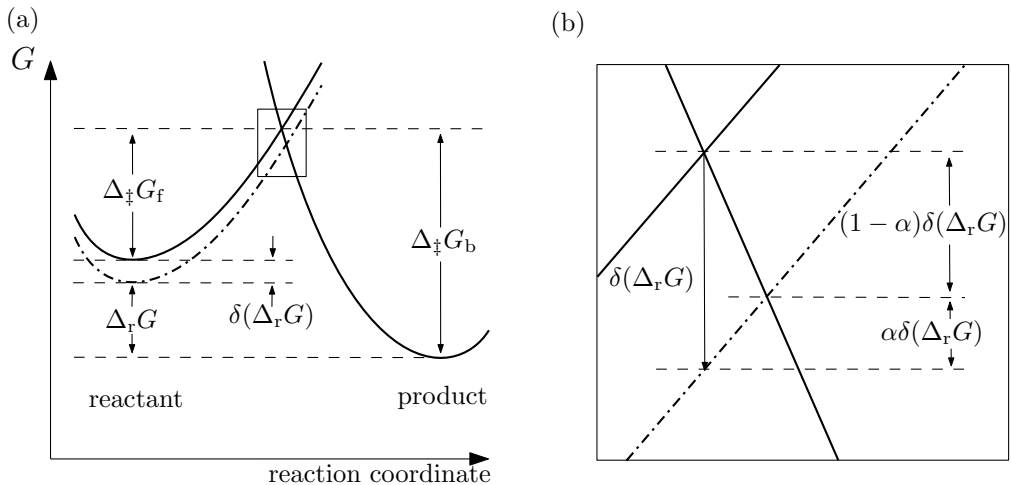

\citet{bell1936theory} and \citet{evans1938inertia} independently derived a principle that relates a change in reaction energy  to a change in the activation energy. Consider the free energy diagram in Fig. \ref{fig:bep}a. Imagine that some parameter (here: the electrode potential) is tuned, causing the reactant curve to shift down, without changing its overall shape. This shift changes the reaction energy by an amount $\delta(\Delta_\mathrm{r} G) > 0$. Note that the reaction energy is negative in this diagram, and the shift makes the reaction energy less negative. By studying the geometrical construction in \autoref{fig:bep}b (similar to the one in \citet{BardFaulkner2001}), we find that the transition state energy increases by $\alpha \delta(\Delta_\mathrm{r} G)$, i.e. 
\begin{equation}
    \Delta_\ddag G_\mathrm{f} \to \Delta_\ddag G_\mathrm{f} + \alpha \delta(\Delta_\mathrm{r} G).
\end{equation}
Conversely, the backward activation energy changes as
\begin{equation}
    \Delta_\ddag G_\mathrm{b} \to \Delta_\ddag G_\mathrm{b} - (1 - \alpha) \delta(\Delta_\mathrm{r} G).
\end{equation}
In the context of electrode kinetics, the parameter $\alpha$ ($0<\alpha<1$) is known as the transfer coefficient \citep{BardFaulkner2001}. 

In this work, the curves of the initial and final states both depend on the electrode potential, through $\phi_0$ and $\phi'$ respectively -- see Fig. \ref{fig:reaction-pe-her}. However, we can still apply the construction shown in Fig. \ref{fig:bep}, as long as we vertically shift the reactant and product curves in Fig. \ref{fig:reaction-pe-her} such that the product curves corresponding to different electrode potentials coincide. Because we only deal with energy differences and not absolute energies, we can still apply the Bell-Evans-Polanyi principle in the same way.

\section{Numerical implementation} \label{app:numerical}
To solve the double-layer model numerically, we first introduce dimensionless quantities. To this end we use an inverse length scale denoted as $\kappa$, 
\begin{equation}
    \kappa = \sqrt{\frac{\beta  e^2}{\varepsilon_\mathrm{w} a^3 }}
\end{equation}
with $\varepsilon_\mathrm{w}\approx78.5\varepsilon_0$ the bulk permittivity of water. Using $\kappa$ we define the dimensionless potential $y_1$, electric field $y_2$, spatial coordinate $\zeta$, rescaled permittivity $\tilde{\varepsilon}$, dimensionless dipole moment magnitude $\tilde{p}$, and dimensionless number density $\chi_i$ as
\begin{align}
    y_1 &= \beta e_0 \phi,  \\
    y_2 &= \frac{\beta e_0 }{\kappa} \frac{\partial \phi}{\partial x}, \\
    \zeta &= \kappa (x-x_2), \\
    \tilde{\varepsilon} &= \frac{\varepsilon}{\varepsilon_\mathrm{w}}, \\
    \tilde{p}&=\frac{\kappa p}{e_0}, \\
    \chi_i &= n_i a^3.
\end{align}
where $p$ is the magnitude of the effective dipole moment of water. 

To fit an effective value for $p$, \citet{booth1951dielectric} considers pure bulk water, where $E=0$. Taking the limit of Equation (8) in the main text as $E\to 0$, the bulk value is found as
\begin{equation}
    \varepsilon_\mathrm{w} = \varepsilon_\infty + \frac13 \beta p^2 n_\mathrm{w}.
\end{equation}
Hence,
\begin{equation}
    p = \sqrt{\frac{3(\varepsilon_\mathrm{w} - \varepsilon_\infty)}{\beta n_\mathrm{w}}},
\end{equation}
where $n_\mathrm{w}=1/a^3=55.5 \text{ M } \times N_\mathrm{A}$. This expression gives $p\approx1.58 \times 10^{-29}\mathrm{~C\;m} = \qty{4.75}{D}$. 

For the numerical solution of the boundary value problem we use SciPy's \texttt{solve\_bvp} function \citep{2020SciPy-NMeth}. To solve Eq. (7) with \texttt{solve\_bvp} we must rewrite the equation as a system of two first-order differential equations. To this end we define
\begin{align}
    F_1 &= -\sum_i z_i \chi_i \\
    F_2 &= - \tilde{p} y_2 \mathcal{L}(\tilde{p}y_2) \chi_\mathrm{w} \sum_i z_i \gamma_i \chi_i \\
    G_1 &= \tilde \varepsilon_\infty \\
    G_2 &= \tilde{p}^2 \chi_\mathrm{w} \mathcal{L}'(\tilde{p}y_2) \\
    G_3 &= \tilde{p}^2 \mathcal{L}^2(\tilde{p} y_2) \chi_\mathrm{w} \left(1-\chi_\mathrm{w}\right)    
\end{align}
where $\mathcal{L}'(x)=\mathrm{d}\mathcal{L}(x)/\mathrm{d} x$, $\mathcal{L}^2(x)=(\mathcal{L}(x))^2$, and $z_i$ the charge number of the various species ($+1$ for cations, $-1$ for anions, $0$ for water). The dimensionless number densities $\chi_i$ are defined in accordance with Eq. (9). The system of first-order differential equations is then
\begin{equation}
    \begin{dcases}
        \frac{\partial y_1}{\partial \zeta} &= y_2 \\
        \frac{\partial y_2}{\partial \zeta} &= \frac{F_1 + F_2}{G_1 + G_2 + G_3}.
    \end{dcases}
\end{equation}
The nondimensionalized boundary conditions are (see Eq. 10)
\begin{equation}
    \begin{dcases}
        y_1(\zeta_\mathrm{end})&=0 \\
        y_1(0) &= \beta e_0 \phi_0 + y_2(0) \kappa x_2,
    \end{dcases}
\end{equation}
where we chose $\zeta_\mathrm{end}$ to correspond to $x= 100 \text{ nm}$, i.e. far away in the electrolyte as compared to the double-layer thickness. As initial $\zeta$-axis we chose a logarithmically spaced axis so that there are more points in the double-layer region and fewer points in the bulk electrolyte. Note that $\phi_0$ is calculated from the electrode potential $\mathsf{E}$ as described in Sec. \ref{sec:potentials}.

To obtain a solution for a metal surface at arbitrary $\phi_0$, we first solve at the pzc ($\phi_0=0$) and then sweep to the desired potential in steps of $0.01$ V, each time using the solution as initial condition for the next iteration. From the solutions $y_1, y_2$, the relevant physical quantities $\phi$, $E$, $c_i=n_i/N_\mathrm{A}$, and $\varepsilon$ were calculated and used as described in the main text.

\begin{table}[t]
\centering
\begin{tabularx}{\linewidth}{@{}XX@{}}
\toprule
\multicolumn{2}{l}{\textbf{General constants}}             \\ \midrule
Boltzmann constant, $k_\mathrm{B}$ & $1.38 \times 10^{-23}$ J/K \\
Elementary charge, $e_0$ & $1.60\times10^{-19}$ C \\
Avogadro's number, $N_\mathrm{A}$ & $6.02 \times 10^{23}$ /mol \\             
Vacuum permittivity, $\varepsilon_0$  & $8.85 \times 10^{-11}$ F/m \\
\midrule
\multicolumn{2}{l}{\textbf{Electrolyte model}  }           \\ 
\midrule
Temperature, $T$ & 298 K \\
Inverse temperature, $\beta$ & $(k_\mathrm{B} T)^{-1}$ \\
Permittivity of pure water \citep{israelachvili2015intermolecular}, $\varepsilon_\mathrm{w}$ & 78.5$\varepsilon_0$  \\
Number density of pure water, $(d_\mathrm{w})^{-3}$ & 55.5 M $\times N_\mathrm{A}$ \\
Lattice spacing, $a$ & $d_\mathrm{w}$ \\
Cation eff. size factor, $\gamma_{+}$ & 4, 5, 6, 7 \\
Anion eff. size factor, $\gamma_{-}$ & 2   \\
Optical permittivity of water \citep{iglivc2019differential}, $\varepsilon_\infty$ & $1.33^2$  \\
Effective dipole moment, $p$ & $\sqrt{\dfrac{3a^3 (\varepsilon_\mathrm{w} - \varepsilon_\infty)}{\beta}}$  \\
Effective ion diameter, $d_\pm$ & $(\gamma_i)^{1/3} a $  \\
Distance of closest approach, $x_2$ & $d_+/2$ \\
Cation bulk concentration, $c_+^*$ & parameter \\
Hydroxide ion concentration, $c_\mathrm{OH^-}^*$  & $10^{-14+\mathrm{pH}}$ \\
\midrule
\multicolumn{2}{l}{\textbf{Metal}}     \\ 
\midrule
pzc of Au(111)  \citep{ojha2020double} \\
$\mathsf E_\mathrm{pzc,Au}$ & \qty{0.51}{V} vs. SHE \\
pzc of Pt(111) \citep{ojha2020double} \\
$\mathsf E_\mathrm{pzc,Pt}$ & \qty{0.3}{V} vs. SHE \\
\midrule   
\end{tabularx}
\caption{Parameter table for the implemented double-layer model.}
\label{tab:parameter-table}
\end{table}

\section{Additional results}
Below we discuss the additional results that are referred to in the main text.

\subsection{Marcus theory} \label{app:marcus}
In the main text we used the Bell-Evans-Polanyi principle to derive an expression for the activation energy; this approach is also referred to as Frumkin-Butler-Volmer theory. Another way to express the activation energy is using Marcus theory \citep{BardFaulkner2001}, which yields
\begin{equation} \label{eq:delta-ddag-g-marcus}
    \Delta_\ddag G = \frac{(\lambda + \Delta_\mathrm{r} G)^2}{4\lambda},
\end{equation}
where $\lambda$ is the reorganization energy. 

We may write the reaction energy in Eq. \eqref{eq:delta-r-g-chempot} in a more convenient form by recognizing the pzc (Eq. \ref{eq:pzc}) and the equilibrium electrode potential of the water reduction sub-step of the Volmer step 
\begin{equation} \label{eq:volmer-equilibrium-pot}
    \mathsf{E}_\mathrm{V1} = \frac{1}{e_0}(\mu_\mathrm{H^*} + \mu_\mathrm{OH^-} - \mu_\mathrm{H_2O})
\end{equation}
where \ce{OH^-} is located in the reaction plane. It follows that
\begin{align} \label{eq:reaction-energy}
    \Delta_\mathrm{r} G = e_0 (\phi_0 - \phi' + (\mathsf{E}_\mathrm{pzc} - \mathsf{E}_\mathrm{V1})).
\end{align}

Inserting $\Delta_\mathrm{r}G$ from Eq. \eqref{eq:reaction-energy} in Eq. \eqref{eq:delta-ddag-g-marcus}, we can express the current density from Eq. \eqref{eq:current-delta-act-g} as
\begin{equation}\label{eq:her-marcus}
    j = - \frac{2 e_0 \bar n_\mathrm{H_2O}}{\beta h}
    \exp \left(-\frac{\beta}{4\lambda} \big(\lambda + e_0(\phi_0 - \phi' + (\mathsf{E}_\mathrm{pzc} - \mathsf{E}_\mathrm{V1}) \big)^2 \right)
\end{equation} 
where $\lambda$ is a free parameter. To evaluate the expression, we also need a value for $\mathsf{E}_\mathrm{V1}$. This value could in principle be extracted from quantum chemical simulations, but does not qualitatively affect the results. Hence, we choose $\mathsf{E}_\mathrm{V1} = \qty{0}{V}$ vs. SHE for simplicity.

\begin{figure*}
    \centering
    \includegraphics[width=\linewidth]{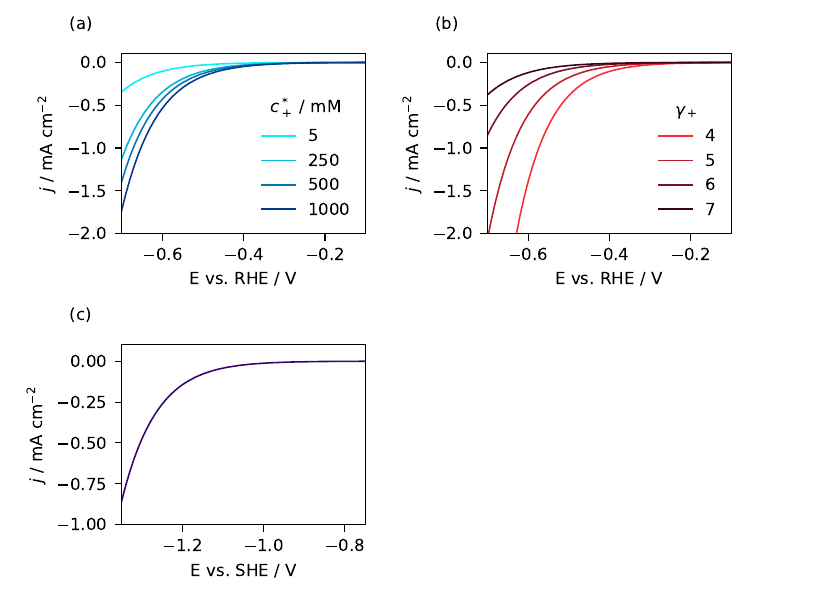}
    \caption{Hydrogen evolution current density on gold, calculated with Eq. \eqref{eq:her-marcus} and setting $\lambda=4.35$ eV. (a) Current density for various cation bulk concentrations $c_+^*$, shown on the RHE scale for pH 11. (b) Current density for various effective cation sizes, shown on the RHE scale for pH 11. (c) pH-independent current density on the SHE scale. Unless indicated otherwise in the legend, $c_+^*=\qty{100}{mM}$, pH 11, and $\gamma_+=6$. }
    \label{fig:gold-marcus}
\end{figure*}

The resulting current density on gold for various concentrations, cation species, and pH is shown in Fig. \ref{fig:gold-marcus}. The experimental conditions are again taken to be the same as in Fig. \ref{fig:gold-experimental}.

\subsection{Quantitative analysis for platinum} 
\begin{figure}
    \centering
    \includegraphics[width=\linewidth]{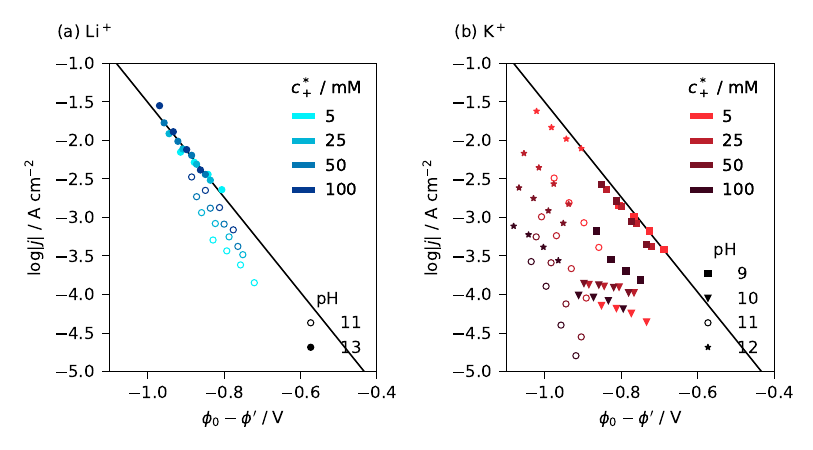}
    \caption{The logarithm of the measured current density on platinum electrodes, $\log |j|$, against the potential drop in the double layer, $\phi_0 - \phi'$, computed from the double-layer model using the corresponding experimental conditions. Data obtained from \citet{monteiro2021understanding}. Subfigure labels indicate the cation species. Parameters: $\gamma_+=7$ for \ce{Li^+}, $5$ for \ce{K^+}. The fit of the \ce{Li^+} data at pH 13 (with $\alpha=0.36)$ is shown as a black line in both subfigures.}
    \label{fig:quantitative-pt}
\end{figure}

The same analysis as in Sec. \ref{sec:quantitative} can be made for the platinum data measured by \citet{monteiro2021understanding}. The result is shown in Fig. \ref{fig:quantitative-pt}. The black line fits the \ce{Li^+} data at pH 13; the fitting value for the transfer coefficient is $\alpha=0.36$.

\label{app:platinum}

\subsection{Quantitative analysis with different parameters} \label{app:ringe}
In the main text we obtained the distance of closest approach $x_2$ from $x_2=d_+/2$, with $d_+$ calculated from Eq. \eqref{eq:gamma}. Here, we consider the case where $x_2$ is specified independently of $\gamma_+$. We use the values given by \citet{ringe2019understanding}, which are based on experiments and a fit of their model to kinetic data. The values of $x_2$ used in the main text and those used by \citeauthor{ringe2019understanding} are compared in Table \ref{tab:x-2}.

\begin{figure}
    \centering
    \includegraphics[width=\linewidth]{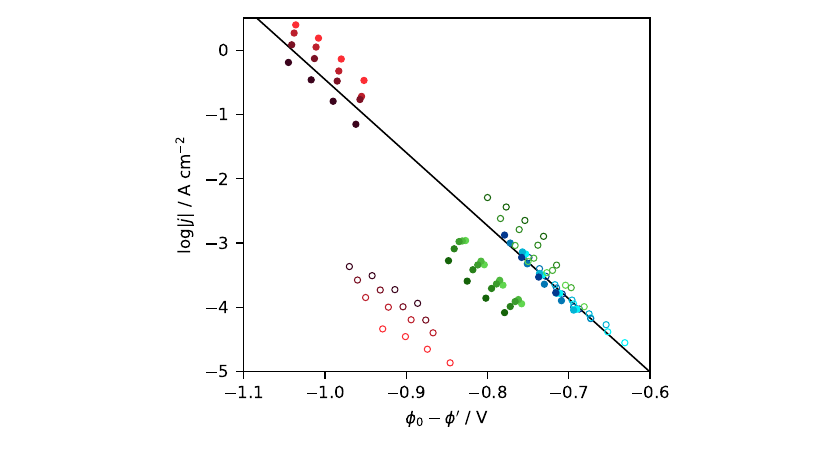}
    \caption{The logarithm of the measured current density on gold electrodes, $\log |j|$, against the potential drop in the double layer, $\phi_0 - \phi'$, computed from the double-layer model using the corresponding experimental conditions and the values of $x_2$ from \citet{ringe2019understanding}. Data obtained from \citet{goyal2021understanding}. For the legend, see Fig. \ref{fig:quantitative}. The black line is a linear fit of the data for \ce{Li^+}, corresponding to $\alpha=0.67$. }
    \label{fig:quantitative-ringe}
\end{figure}

Figure \ref{fig:quantitative-ringe} shows the results analogous to those in Fig. \ref{fig:quantitative}, but now using the new values of $x_2$ to calculate $\phi_0 - \phi'$. The data is visualized according to the same legend as in Fig. \ref{fig:quantitative}.

The data for \ce{Li^+} still falls on one line with $\alpha=0.67$. The most important difference between the results in the main text and the result in Fig. \ref{fig:quantitative-ringe} is that here the data for \ce{K^+} at pH $13$ lies more in line with the data for \ce{Li^+}. However, as remarked in the main text, the different concentrations still do not fall on the same line. 

We do not use the values of $x_2$ from \citeauthor{ringe2019understanding} in the results presented in the main text for the following reasons. First, this choice reduces the number of parameters in our model. Second, the large differences between the values of $x_2$ of \citeauthor{ringe2019understanding} leads to a much larger difference in $\phi_0 - \phi'$ for various effective ion sizes. As a result, the difference in current between various cation species is much larger than what is observed in the experimental data (Fig. \ref{fig:gold-species-trend}). Finally, the fact that only one additional data set follows the expected trend is not enough evidence to conclude that this set of parameters is more realistic. 

\begin{table}
\begin{tabularx}{\linewidth}{@{}XXX@{}}
\toprule
Ion & $x_2$ (this work) & $x_2$ (from \citep{ringe2019understanding}) \\ \midrule
\ce{Cs^+} & 2.46    & 3.5      \\
\ce{K+}  & 2.65    & 4.1      \\
\ce{Na+} & 2.82    & 5.2      \\
\ce{Li+} & 2.97    & 5.8      \\ \bottomrule
\end{tabularx}
\caption{Comparison of the distance of closest approach $x_2$ for various ion species used in this work and the work of \citet{ringe2019understanding}.}
\label{tab:x-2}
\end{table}

\subsection{Effect of the choice of pzc} \label{app:pzc}
As described in Sec. \ref{sec:potentials} in the main text, the choice of pzc in our model is rather arbitrary. In this section, we show several results computed for different values of the pzc. The trends are qualitatively the same, and so the discussion of the results in the main text is not affected by these results.

\begin{figure}
    \centering
    \includegraphics[width=\linewidth]{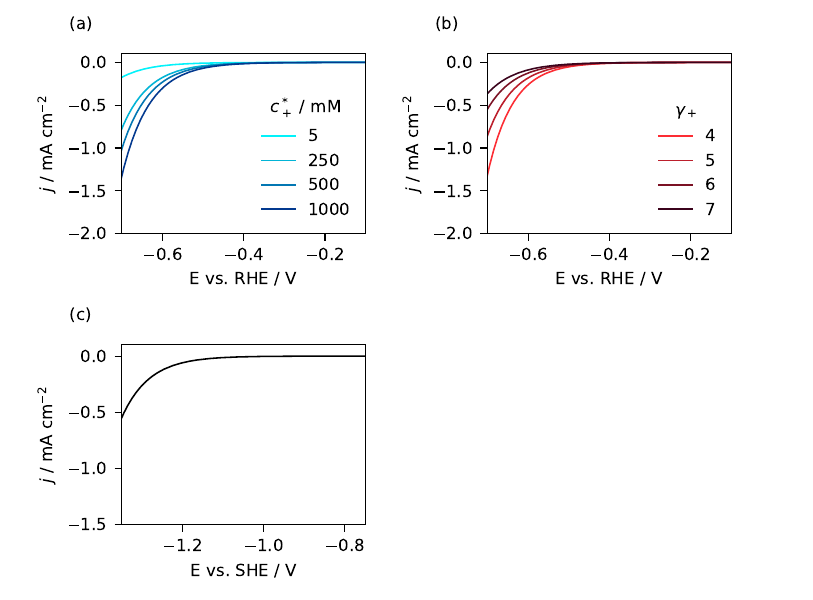}
    \caption{Results of Figure \ref{fig:gold-fbv} recomputed with $\mathsf{E}_\mathrm{pzc}=\qty{-0.54}{V}$ vs. SHE and $(\Delta_\ddag G)_\mathrm{pzc}=1$ eV.} 
    \label{fig:gold-fbv-pzc-054}
\end{figure}

Figure \ref{fig:gold-fbv-pzc-054} shows the results in case the pzc of gold also shifts to more negative values in alkaline media. We take the most negative value mentioned in Sec. \ref{sec:potentials}, $\mathsf{E}_\mathrm{pzc}=\qty{-0.54}{V}$ vs. SHE. To rescale the current to the same order of magnitude as the other results, we set $(\Delta_\ddag G)_\mathrm{pzc}=1$ eV.

We also recomputed the results of Figure \ref{fig:pt-current} with this more negative value of the pzc, $\mathsf{E}_\mathrm{pzc}=\qty{-0.54}{V}$ vs. SHE. It was not necessary to adjust the value of $C$. Although the curves show a different shape, the trends remain qualitatively the same.

\begin{figure}
    \centering
    \includegraphics[width=\linewidth]{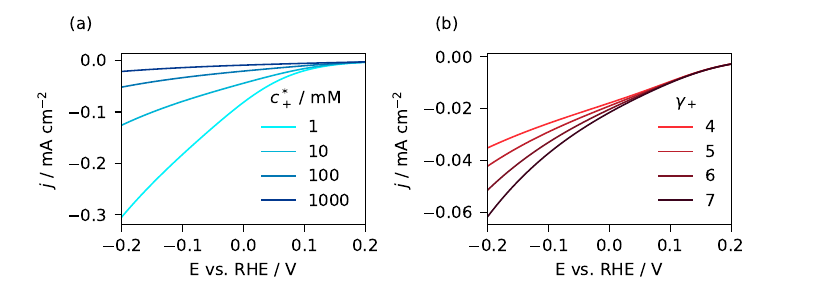}
    \caption{Results of Figure \ref{fig:pt-current} recomputed with $\mathsf{E}_\mathrm{pzc}=\qty{-0.54}{V}$ vs. SHE.}
    \label{fig:pt-pzc-054}
\end{figure}

\begin{figure}
    \centering
    \includegraphics[width=\linewidth]{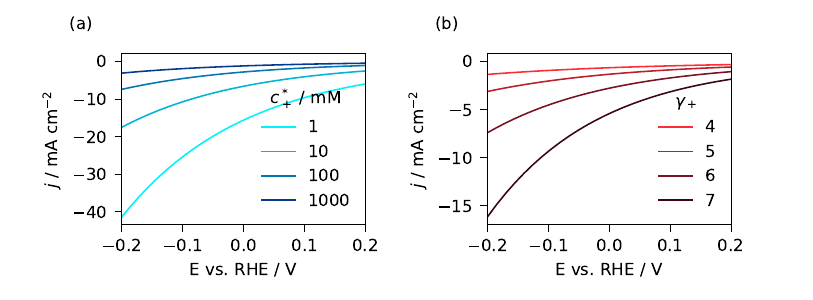}
    \caption{Results of Figure \ref{fig:pt-current} recomputed with $\mathsf{E}_\mathrm{pzc}=\qty{0.51}{V}$ vs. SHE.}
    \label{fig:pt-pzc-051}
\end{figure}

To stress that the different trends in platinum and gold are due to the different rate-limiting step, and not due to a difference in pzc, we also recompute the \ce{OH^-}-transfer-limited current using a pzc of $\mathsf{E}_\mathrm{pzc}=\qty{0.51}{V}$, the pzc that we assumed for gold. These results are shown in Fig. \ref{fig:pt-pzc-051}. The trends are the same as in Fig. \ref{fig:pt-current} and Fig. \ref{fig:pt-pzc-054}, again showing that the trends do not depend on the pzc.

\begin{figure}
    \centering
    \includegraphics[width=\linewidth]{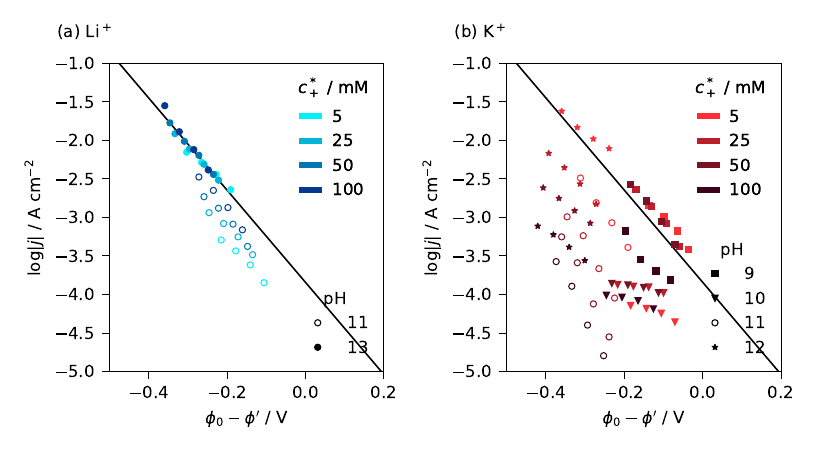}
    \caption{Results of Fig. \ref{fig:quantitative-pt} recomputed using $\mathsf{E}_\mathrm{pzc}=\qty{-0.54}{V}$. }
    \label{fig:quantitative-pt-pzc-054}
\end{figure}

Finally, we recompute the results of Fig. \ref{fig:quantitative-pt} with $\mathsf{E}_\mathrm{pzc}=\qty{-0.54}{V}$, as shown in Fig. \ref{fig:quantitative-pt-pzc-054}. The results are not strongly affected by the choice of pzc, suggesting that the quantity $\phi_0-\phi'$ scales with the bulk cation concentration in the same way regardless of the choice of pzc. 

\clearpage

\bibliographystyle{elsarticle-num-names} 
\bibliography{bibliography}

\end{document}